\documentclass[12pt,dvips,a4paper]{article}
\usepackage[pdftex]{graphicx}
\usepackage{amssymb}
\usepackage{amsmath}
\usepackage{setspace}
\usepackage[round]{natbib}

\begin{document}

\renewcommand{\title}{Meso-scale modelling of the size effect on the fracture process zone of concrete}

\begin{center} \begin{LARGE} \textbf{\title} \end{LARGE} \end{center}

\begin{center}
P. Grassl$^{1*}$, David Gr\'{e}goire$^2$, Laura Rojas Solano$^2$ and Gilles Pijaudier-Cabot$^2$\\
$^1$ School of Engineering, University of Glasgow, United Kingdom\\
$^2$ Laboratoire des fluides complexes et leurs r\'{e}servoirs, Universit\'{e} de Pau et des Pays de L'Adour, France
\end{center}

$^*$~corresponding author:\\
Email: peter.grassl@glasgow.ac.uk\\
Phone: +44 141 330 5208\\
Fax: +44 141 330 4557

\section*{Abstract}
The size effect on the fracture process zone in notched and unnotched three point bending tests of concrete beams is analysed by a meso-scale approach.
Concrete is modelled at the meso-scale as stiff aggregates embedded in a soft matrix separated by weak interfaces.
The mechanical response of the three phases is modelled by a discrete lattice approach.
The model parameters were chosen so that the global model response in the form of load-crack mouth opening displacement curves were in agreement with experimental results reported in the literature.
The fracture process zone of concrete is determined numerically by evaluating the average of spatial distribution of dissipated energy densities of random meso-scale analyses.
The influence of size and boundary conditions on the fracture process zone in concrete is investigated by comparing the results for beams of different sizes and boundary conditions.
    
Keywords: Size effect; Fracture process zone; Lattice; Concrete; Meso-scale; Fracture

\section{Introduction}
A major question for extrapolating the results of small-scale laboratory tests to large-scale applications is size effect.
For plain concrete structures subjected to bending, the nominal strength depends strongly on the size of the structure. 
Here, size effect is the dependence of the dimensionless nominal strength of a beam on its depth, when geometrically similar structures are compared. 
The smaller the structure, the greater is the nominal strength.
Experimental results show that this size effect follows neither the strength limit nor linear elastic fracture mechanics \citep{Baz01}.
Currently, three main theories for describing the size effect are promoted in the literature, namely statistical theories of random strength, stress-redistribution and energy release, and theory of crack fractality \citep{Baz01}.
There is no full consensus on the range of applicability of these theories for concrete structures of different geometries and subjected to different loading, since the results of several of these theories are similar for the size range of experimental results available \citep{BazYav05}.
Most experimental studies are only able to demonstrate the effect on the strength for a small size range, since large scale experiments are too difficult to carry out.
Furthermore, in most experiments, only global quantities, such as the overall strength of the structure, are determined.
If local quantities, such as the spatial distribution of dissipated energy and fracture patterns could be obtained, our understanding of fracture processes would improve, which might reduce the ambiguity in the modelling of the size effect on nominal strength.
Experimentally, local results have been obtained by acoustic emission measurements \citep{Lan99,OtsDat00,HaiPijDub04} and X-ray tomography \citep{LanNagKea03}. 
Digital image correlation techniques have been applied to study fracture processes in concrete \citep{ChoSha97} and other materials \citep{GreLohJus11,EspJusLat11}.
However, more detailed results are required to fully understand the mechanics of the evolution of fracture processes of concrete, especially for different sizes.

The objective of this work is to contribute to the understanding of fracture processes by investigating numerically the size effect on fracture process zones in concrete structures.
Here, fracture process zone is defined as the zone in which energy is dissipated at a certain stage of analysis.
Three point bending beams of different sizes and geometries were simulated using a meso-scale model, in which aggregates embedded in a matrix separated by weak interfacial transition zones were discretised by a network of lattice elements.
The input parameters of the model were chosen so that the global model response in the form of load-crack mouth opening displacement (CMOD) curves were in agreement with recently obtained experimental results reported by \citet{GreSolPij11}.
The fracture processes were evaluated by computing the rate of dissipated energy of individual lattice elements.
The fracture process zone was determined from the average of spatial distributions of dissipated energy densities of multiple analyses with randomly arranged aggregates following the same aggregate distributions and random fields of material properties.
This strategy for determining the fracture process zone by averaging the spatially distributed rate of dissipation has been developed recently by \citet{JirGra07} and \citet{GraJir10}.
The contribution of the present study is to apply this strategy to the meso-scale modelling of the size effect on the fracture process zone, which should provide a better insight into fracture processes and, therewith, a better description of size effect.

Modelling approaches for fracture in concrete are divided in three groups. 
Continuum approaches describe the fracture process by higher-order constitutive models, such as integral-type nonlocal models \citep{BazJir02}.
In continuum models with discontinuities, cracks are described as displacement discontinuities, which are embedded into the continuum description \citep{JirZim01}.
Finally, discrete approaches describe the nonlinear fracture process as failure of discrete elements, such as lattices of trusses and beams \citep{Kaw78,Cun79}.
In lattice approaches, the connectivity between nodes is not changed so that contact determination is simplified.
Lattice models are mainly suitable for analyses involving small strains \citep{HerHanRou89,SchMie92b,BolSai98}.
In recent years, one discrete approach based on a lattice determined by Voronoi tesselation has been shown to be suitable for fracture simulations \citep{BolSai98}.
This lattice approach is robust, computationally efficient, and allows for fracture description by a stress-inelastic displacement relationship, similar to continuum models with discontinuities. Therefore, this approach is suitable for modelling interfaces at the meso-scale of concrete. 
Furthermore, with a specially designed constitutive model, the crack openings obtained with this approach were shown to be independent of the size of the elements \citep{BolSuk05}. 
This type of model is used in the present study.
In the lattice framework, the meso-structure of concrete is either idealised by modelling the interaction of two aggregates by a single element \citep{ZubBaz87} or by mapping the meso-structure of concrete on the lattice by employing a field of spatially varying material properties \citep{SchMie92b}.
The latter approach is chosen in the present study, since the finer resolution allows for a detailed description of the tortuosity of crack patterns, which is important for the modelling of the fracture process zone \citep{GraJir10}.
The constitutive response for the individual phases can either be described by micro-mechanics or phenomenological constitutive models, commonly based on the theory of plasticity, damage mechanics or a combination of the two.
For predominantly tensile loading, an isotropic damage model has shown to provide satisfactory results \citep{GraJir10}. Such a model is used here.

The present work is based on several assumptions.
In the chosen idealisation of the meso-structure only large aggregates were considered, which are embedded in a mortar matrix separated by interfacial transition zones.
The aggregates were assumed to be linear elastic and stiffer than the matrix, whereas the interfacial transition zone was assumed to be weaker and more brittle than the matrix.
The material constants for the constitutive models of the three phases were chosen by comparing the global results of analyses and experiments assuming certain ratios of the properties of different phases.
For instance, aggregates were assumed to be twice as stiff as the matrix, which in turn is twice as strong and ductile as the interfacial transition zone. 
These chosen ratios are supported by experimental results reported in the literature \citep{HsuSla63, Mie97}.
Furthermore, the present study was limited to two-dimensional plane stress analyses with aggregates idealised as cylindrical inclusions.

\section{Meso-scale modelling approach}

\begin{figure}
\begin{center}
\begin{tabular}{cc}
\includegraphics{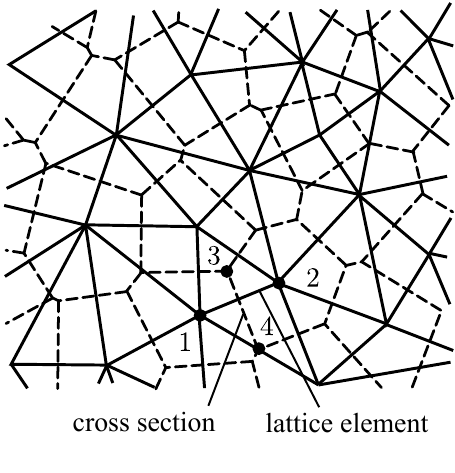} & \includegraphics{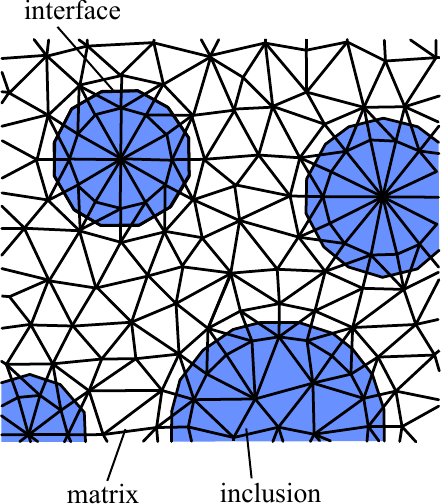}\\
(a) & (b) \vspace{1cm}\\
\includegraphics{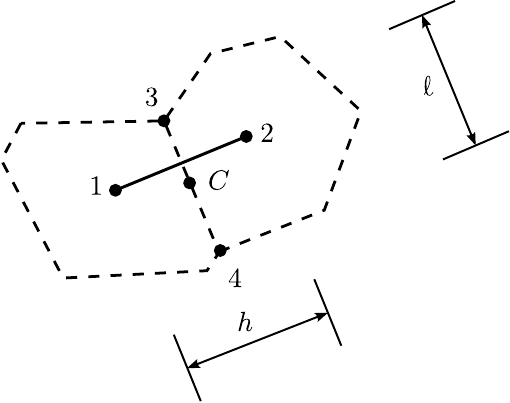} & \includegraphics{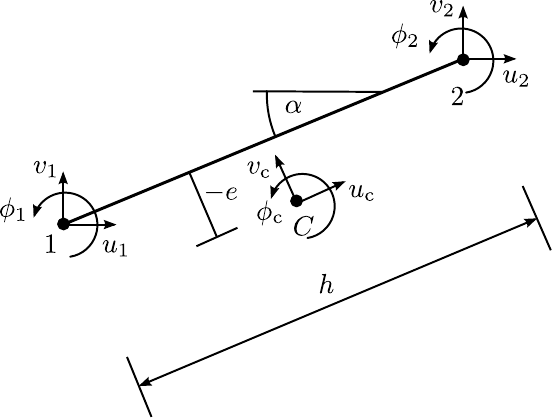}\\
(c) & (d)
\end{tabular}
\end{center}
\caption{(a) Set of lattice elements (solid lines) with middle cross-sections (dashed lines) obtained from the Voronoi tesselation of the domain, (b) arrangement of lattice elements around inclusions. (c) and (d) Lattice element in the global coordinate system.}
\label{fig:lattice}
\end{figure}

The present 2D plane stress modelling approach for fracture in concrete is based on a lattice formed by discrete structural elements \citep{GraJir10}.
The nodes of the lattice are randomly located in the domain, subject to the constraint of a minimum distance $d_{\rm min}$.
The lattice elements are obtained from the edges of the triangles of the Delaunay triangulation of the domain (solid lines in Fig.~\ref{fig:lattice}a), whereby the middle cross-sections of the lattice elements are the edges of the polygons of the dual Voronoi tesselation (dashed lines in Fig.~\ref{fig:lattice}a).  
For the discretely modelled aggregates, the lattice nodes are placed at special locations, such that the middle cross-sections of the lattice elements form the boundaries between aggregates and mortar \citep{BolBer04} (Fig.\ref{fig:lattice}b).

Each lattice node possesses three degrees of freedom, namely two translations and one rotation. 
In the global coordinate system, shown in Fig.~\ref{fig:lattice}c~and~d, the degrees of freedom $\mathbf{u}_{\rm e} = (u_1, v_1, \phi_1, u_2, v_2, \phi_2)^{\rm T}$ of the lattice nodes are linked to three displacement discontinuities $\mathbf{u}_{\rm c} = (u_{\rm c}, v_{\rm c}, \phi_{\rm c})^{\rm T}$ in the local co-ordinate system at point $C$, which is located at the centre of the middle cross-section of the element. 
The distance $e$ between the center $C$ and the midpoint of the lattice element is defined to be positive if it is located on the left of the element.
The relation between the degrees of freedom and the displacement discontinuities at $C$ is 
\begin{equation}\label{eq1}
\mathbf{u}_{\rm c} = \mathbf{B} \mathbf{u}_{\rm e}
\end{equation}
where
\begin{equation} \label{eq:bMatrix}
\mathbf{B} = \begin{bmatrix}
-\cos \alpha & - \sin \alpha & -e & \cos \alpha & \sin \alpha & e\\
\sin \alpha & - \cos \alpha & -h/2 & \cos \alpha & \sin \alpha & -h/2\\
0 & 0 & \sqrt{I/A} & 0 & 0 & -\sqrt{I/A} 
\end{bmatrix}
\end{equation}
Here, the cross-sectional area is defined as $A = \ell \times t$, where $\ell$ is the width of the middle cross-section and $t$ is the out-of-plane thickness.
Accordingly, the second moment of area is $I = \ell^3t/12$.

The displacement discontinuities $\mathbf{u}_{\rm c}$ at point $C$ are transformed into strains $\boldsymbol{\varepsilon}  = \mathbf{u}_{\rm c} /h = (\varepsilon_{\rm n}, \varepsilon_{\rm s}, \varepsilon_{\rm \phi})^{T}$, where $h$ is the distance between the two lattice nodes.
The strains are related to the stresses $\boldsymbol{\sigma} = (\sigma_{\rm n}, \sigma_{\rm s}, \sigma_{\rm \phi})^{T}$ by an isotropic damage model.
The subscripts $n$ and $s$ refer to the normal and shear components of the strain and stress vector.
The stiffness matrix of the lattice element has the form
\begin{equation}
\mathbf{K} = \dfrac{A}{h} \mathbf{B}^{\rm T} \mathbf{D} \mathbf{B}
\end{equation}
where $\mathbf{D}$ is the material stiffness matrix.

Heterogeneous materials are characterised by spatially varying material properties.
In the present work this is reflected at two levels.
Aggregates with diameters greater than $\phi_{\rm min}$  are modelled directly.
The random distribution of the aggregate diameters $\phi$ is defined by the cumulative distribution function \citep{GraRem08}. 
The aggregates are placed randomly within the domain to be analysed, avoiding overlap of aggregates. 
The heterogeneity represented by finer particles is described by autocorrelated random fields of tensile strength and fracture energy, which are assumed to be fully correlated. 
The random fields are characterised by an autocorrelation length that is independent of the spacing of lattice nodes \citep{GraBaz09}.

An isotropic damage model is used to describe the constitutive response of ITZ and mortar.
In the following section, the main equations of the constitutive model are presented. 
The stress-strain law reads
\begin{equation} \label{eq:totStressStrain}
\boldsymbol{\sigma} = \left(1-\omega \right) \mathbf{D}_{\rm e} \boldsymbol{\varepsilon}  = \left(1-\omega\right) \bar{\boldsymbol{\sigma}}
\end{equation}
where $\omega$ is the damage variable, 
$\mathbf{D}_{\rm e}$ is the elastic stiffness and $\bar{\boldsymbol{\sigma}} = \left(\bar{\sigma}_{\rm n}, \bar{\sigma}_{\rm s}, \bar{\sigma}_{\rm \phi}\right)^T$ is the effective stress.

The elastic stiffness 
\begin{equation}\label{eq:elasticStiffness}
\mathbf{D}_{\rm e} = \begin{bmatrix} E & 0 & 0\\
0 & \gamma E & 0\\
0 & 0 & E
\end{bmatrix}
\end{equation}
depends on model parameters
$E$ and $\gamma$, which  control Young's modulus and Poisson's ratio of the equivalent continuum \citep{GriMus01}.
Equations~(\ref{eq:bMatrix})~and~(\ref{eq:elasticStiffness}) were chosen so that for $h=\ell$ and $e=0$ the stiffness matrix $\mathbf{K}$ reduces to the Bernoulli beam stiffness matrix (\citet{BolSai98}).
For a regular triangular lattice and plane stress, the continuum Poisson's ratio is
\begin{equation}\label{eq:Poisson}
\nu_{\rm c} = \dfrac{1-\gamma}{3+\gamma}
\end{equation}
and the continuum Young's modulus is
\begin{equation}\label{eq:Youngs}
E_{\rm mc} = 2 E \left(\dfrac{1+\gamma}{3+\gamma}\right)
\end{equation}

For a positive shear stiffness, i.e.\ $\gamma >0$, Poisson's ratio is limited to $\nu_{\rm c}<1/3$, which is acceptable for concrete but may be unrealistic for certain other materials. 
For the irregular lattice used in the present study, the expressions in (\ref{eq:Youngs})~and~(\ref{eq:Poisson}) are only approximate. 
The damage parameter $\omega$ is a function of a history variable $\kappa$, which is determined by the loading function
\begin{equation}
f(\boldsymbol{\varepsilon},\kappa) = \varepsilon_{\rm eq} \left( \boldsymbol{\varepsilon} \right) - \kappa 
\end{equation}
and the loading-unloading conditions 
\begin{equation}\label{loadunload}
f \leq 0 \mbox{,} \hspace{0.5cm} \dot{\kappa} \geq 0 \mbox{,} \hspace{0.5cm} \dot{\kappa} f = 0
\end{equation}

Here, the equivalent strain is defined as
\begin{equation} \label{eq:equiv}
\varepsilon_{\rm eq}(\varepsilon_{\rm n},\varepsilon_{\rm s}) = \dfrac{1}{2} \varepsilon_0 \left( 1-c \right) + \sqrt{\left( \dfrac{1}{2} \varepsilon_0 (c-1) + \varepsilon_{\rm n}\right)^2 + \dfrac{ c \gamma^2 \varepsilon_{\rm s}^2}{q^2}}
\end{equation} 
where $\varepsilon_0$, $c$ and $q$ are model parameters, which are directly related to the strength and stiffness of the equivalent continuum of the lattice elements.
The present equivalent strain definition depends only on the first two strain components. However, all three effective stress components in (\ref{eq:totStressStrain}) are reduced by the damage parameter $\omega$.
\begin{figure}
\begin{center}
\begin{tabular}{cc}
\includegraphics[width=7cm]{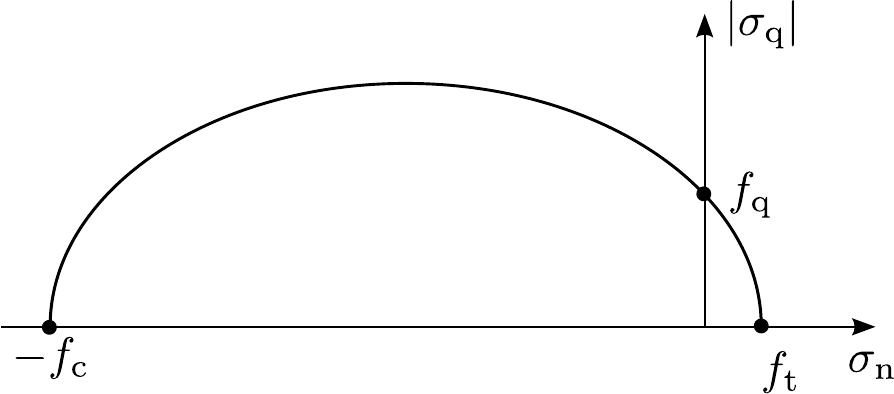} & \includegraphics[width=5cm]{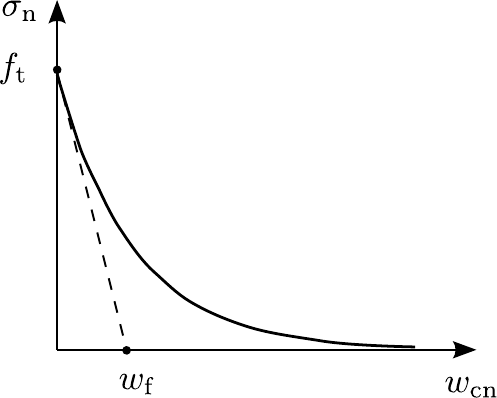}\\
(a) & (b)
\end{tabular}
\caption{(a) Elliptic strength envelope in the nominal stress space and (b) exponential stress crack opening curve.}
\label{fig:envelope}
\end{center}
\end{figure}
This equivalent strain definition results in an elliptic strength envelope shown in Fig.~\ref{fig:envelope}.
For pure tension, the nominal stress is limited by the tensile strength $f_{\rm t} = E \varepsilon_{0}$.
For pure shear and pure compression, the nominal stress is limited by the shear strength $f_{\rm q} = q f_{\rm t}$ and the compressive strength $f_{\rm c} = c f_{\rm t}$, respectively.

The expression for the damage parameter $\omega$ is derived by considering the case of pure tension. 
The softening curve of the stress-strain response in pure tension is chosen as
\begin{equation} \label{eq:exp}
\sigma_{\rm n} = f_{\rm t} \exp \left(-\dfrac{w_{\rm cn}}{w_{\rm f}}\right)
\end{equation}
where $w_{\rm cn} = \omega h \varepsilon_{\rm n}$ is considered as a crack opening in monotonic pure tension (Fig.~\ref{fig:envelope}b). 
The stress-strain relation in pure tension can also be expressed by (\ref{eq:totStressStrain}) in terms of the damage variable as
\begin{equation}\label{eq:uni}
\sigma_{\rm n} =  \left(1-\omega \right) E \varepsilon_{\rm n}
\end{equation}
Comparing the right-hand sides of (\ref{eq:exp}) and (\ref{eq:uni}), 
and replacing $\varepsilon_{\rm n}$ by $\kappa$, we obtain the equation
\begin{equation}
\left(1-\omega \right) \kappa = \varepsilon_0\exp \left(-\dfrac{\omega h \kappa}{w_{\rm f}}\right)
\end{equation}
from which the damage parameter $\omega$ can be determined iteratively using, for instance, a Newton-Raphson method.
Parameter $w_{\rm f}$ determines the initial slope of the softening curve and is related to the meso-level fracture energy $G_{\rm f} = f_{\rm t} w_{\rm f}$, which corresponds to the total area under the stress crack opening curve in Fig.~\ref{fig:envelope}b.
The crack openings obtained with this constitutive model are independent of the length of the element, which has been shown for a very similar constitutive model in \citet{GraJir10}.

\section{Analysis of three point bending tests}

\begin{figure}
\begin{center}
\begin{tabular}{cc}
\includegraphics[width=8cm]{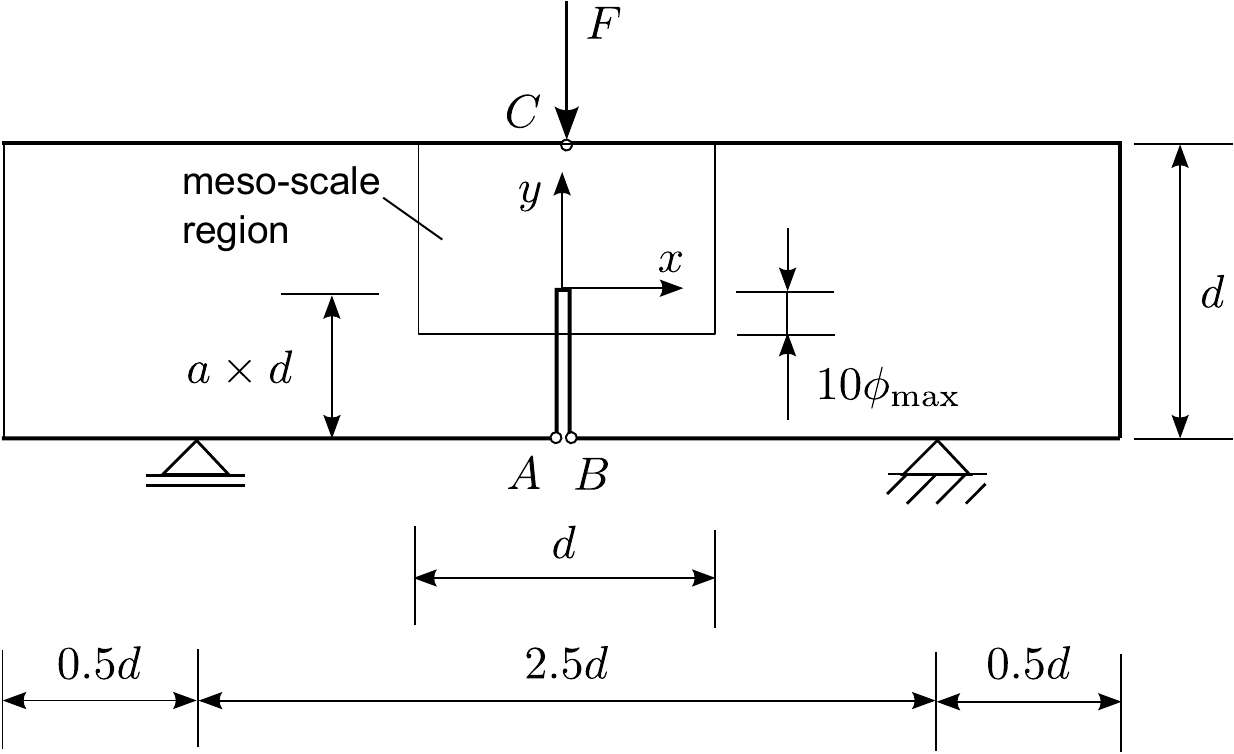} & \includegraphics[width=8cm]{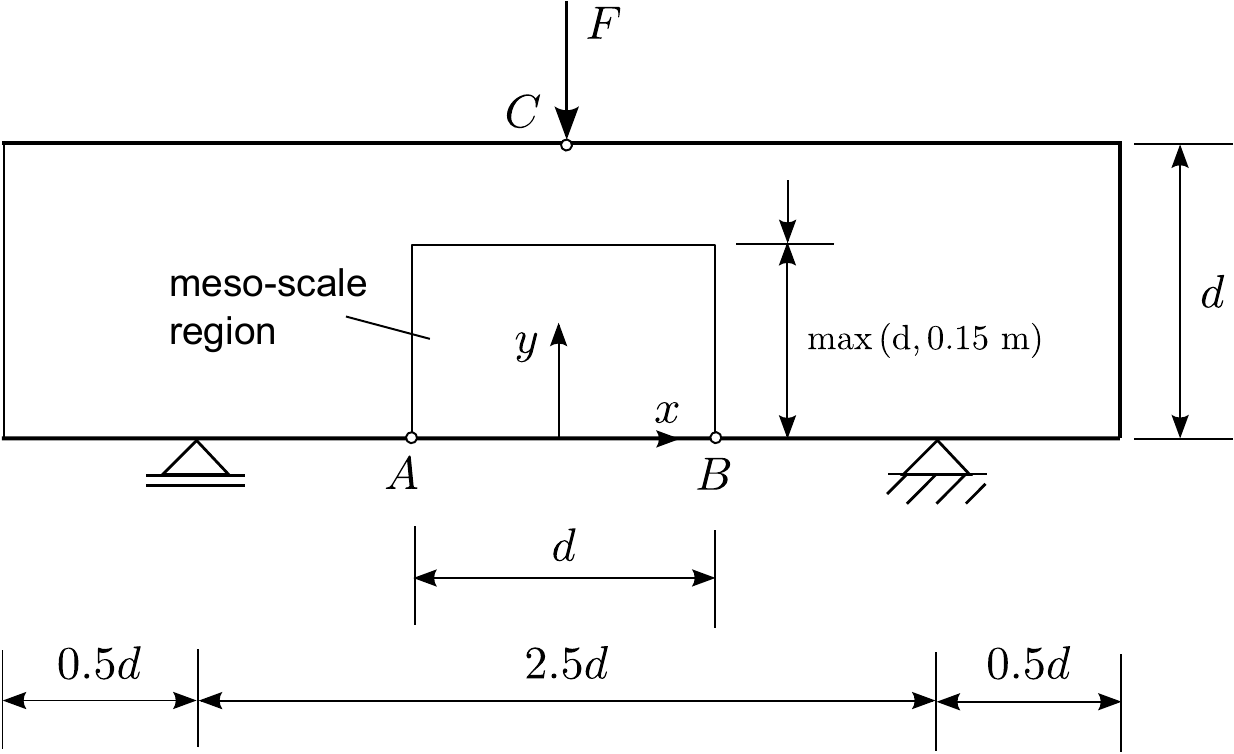}\\
(a) & (b)
\end{tabular}
\caption{Geometries of three-point bending tests for (a) notched and (b) unnotched beams.}
\label{fig:geometry}
\end{center}
\end{figure}

In the present section the geometry, loading setup and results of meso-scale analyses of notched and unnotched three-point bending tests are described.
The numerical results are compared to experiments reported in \citet{GreSolPij11}.
A schematic drawing of the geometries of notched and unnotched beams is shown in Fig.~\ref{fig:geometry}a~and~b.
Four beam depths, $d = 50$, $100$, $200$ and $400$~mm, and three notch lengths, $a = 0$, $0.2$ and $0.5$, were considered.
For the different sizes, all dimensions were scaled proportionally to the beam depth, except the out-of-plane thickness, which was kept constant for all sizes at $t=0.05$~m, see \citet{GreSolPij11}.
In the analyses, a zero notch thickness was assumed, which idealises the experimental procedure in which a thin metal plate of constant thickness was used to cast the notch.
The load and support reactions were applied by support plates of a width of $0.2d$.
Only the middle parts of the beams were modelled using the meso-scale approach (Figs.~\ref{fig:geometry}a~and~b). 
The remaining parts of the beam were modelled by elastic properties describing the average of the elastic response of matrix, aggregate and interfacial transition zone of the composite in the meso-scale region.
The aggregate volume fraction was chosen as $\rho=0.3$ with a maximum and minimum aggregate diameter of $\phi_{\rm max}=10$~mm and $\phi_{\rm min}=5$~mm, respectively. The approach to generate the distribution of aggregate diameters is described in \citet{GraRem08}.
Furthermore, the random field was characterised by the autocorrelation length $l_{\rm a} = 1$~mm and the coefficient of variation $c_{\rm v} = 0.2$.
For more information on the generation of random fields and the definition of $l_{\rm a}$, see \citet{GraBaz09}.
For the random placement of the aggregates, a trial and error approach is chosen so that overlap of aggregates is avoided. Overlap with specimen boundaries and the notch are allowed.

The model constants for the three phases were chosen according to Tab.~\ref{tab:param}, where parameters $\bar{f}_{\rm t}$ and $\bar{G}_{\rm f}$ in Tab.~\ref{tab:param} are the mean values of the random fields of tensile strength $f_{\rm t}$ and fracture energy $G_{\rm f}$, respectively. 
\begin{table}
\begin{center}
\caption{Model parameters.}
\label{tab:param}
\begin{tabular}{cccccccc}
\hline \\
 & $E$ [GPa] & $\gamma$ & $\bar{f}_{\rm t}$~[MPa] & $q$ & $c$ & $\bar{G}_{\rm f}$~[N/m]\\
Matrix & $44$ & $0.33$ & $3.8$ & $2$ & $10$ & $86$\\
Interface & $58.7$ & $0.33$ & $1.9$ & $2$ & $10$  & $43$\\
Aggregate & $88$ & $0.33$ & - & - & - & -\\
Average & $53$ & $0.33$ & - & - & - & -\\
\hline\\
\end{tabular}
\end{center}
\end{table}

These values were chosen so that the global model results in form of load-CMOD curves for different beam sizes and boundary conditions were in agreement with experimental results reported in \citet{GreSolPij11}.
However, not all parameters were varied independently of each other to obtain this agreement.
Instead, several constraints were applied, motivated by experimental and numerical results reported in the literature.
Firstly, the ratio of the stiffnesses for aggregate and matrix was kept constant and equal to two. 
Furthermore, the tensile strength of matrix was assumed to be twice of the strength of the interfacial transition. These ratios are in the range of experimental results reported in the literature \citep{HsuSla63, Mie97}.
Secondly, lattice elements crossing the boundary between aggregates and cement paste represent the average response of the interfacial transition zones and the two adjacent materials \citep{GraJir10}.
In all the analyses, the length of the lattice elements is significantly greater than the width of the interfacial transition zone. 
Therefore, the Young's modulus of the interface elements is 
\begin{equation}
E_{\rm{I}} = \dfrac{2}{\dfrac{1}{E_{\rm m}} + \dfrac{1}{E_{\rm a}}}
\end{equation}
where $E_{\rm m}$ and $E_{\rm a}$ are the Young's moduli of matrix and aggregate, respectively.
Furthermore, the model parameters for the elastic response outside the meso-scale region were chosen so that the response represents the average elastic behaviour of the meso-scale region.
Consequently, the peak and post-peak response was mainly controlled by two free parameters, namely the tensile strength $f_{\rm t,m}$ and fracture energy $G_{\rm f,m}$ of the matrix.  

The analyses were controlled by the crack mouth opening displacement CMOD, which is the relative horizontal displacement of the points $A$ and $B$ shown in Fig.~\ref{fig:geometry}. For the notched specimens ($a=0.2$ and $0.5$), the points were located at the end of the notch. For the unnotched specimen ($a=0$), the two points were apart a distance $d$, since the location of the fracture process zone initiating from the surface was indeterminate.
The CMOD was increased incrementally up to complete failure of the specimen, which in the present analyses was defined as a residual load of less than 1/100 of the peak value. For the unnotched specimen, for which the region in which the meso-scale model is applied is smaller than the depth of the beam ($d = 200$ and $400$~mm), the analyses were stopped shortly after the peak of the load-CMOD curve.

\begin{figure}
\begin{center}
\includegraphics[width=10cm]{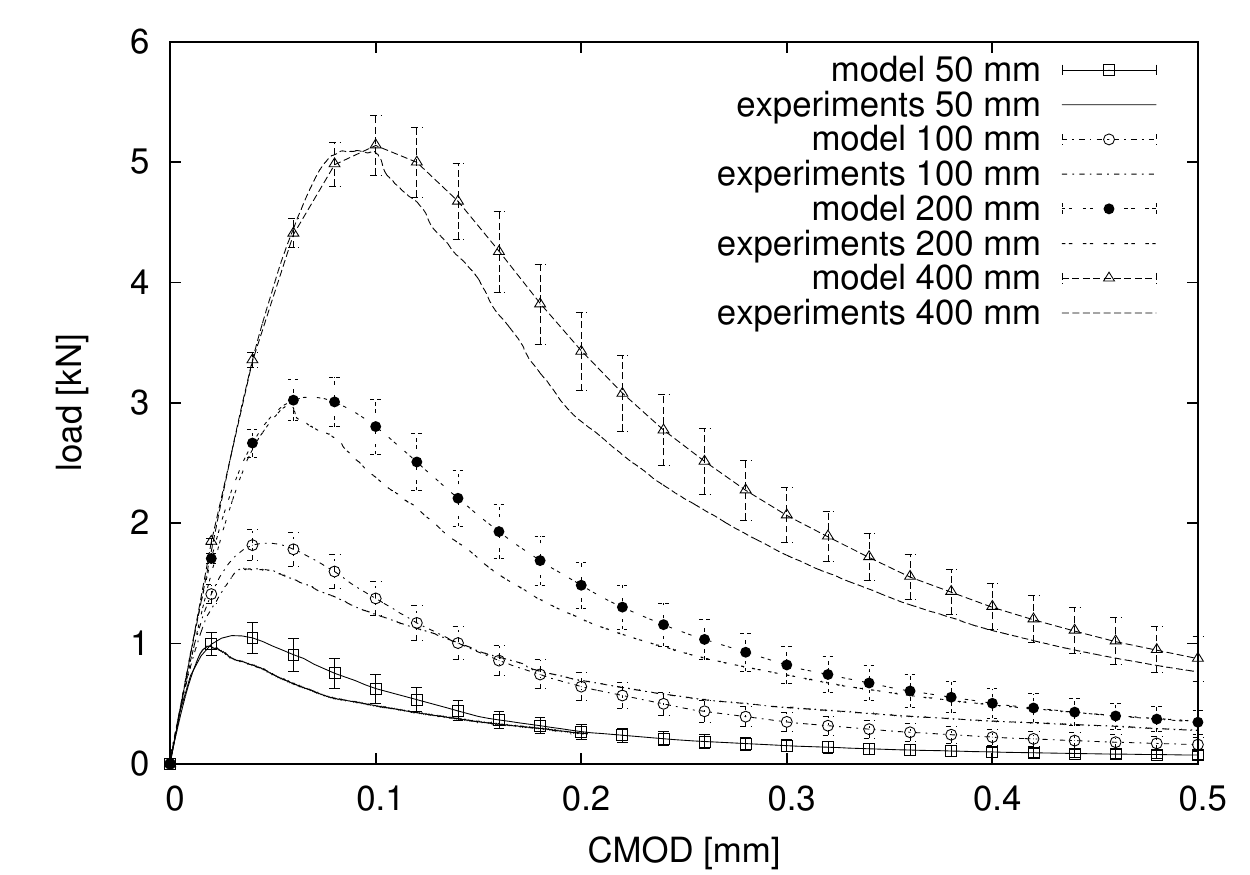}
\end{center}
\caption{Comparison of load versus CMOD of analyses and experiments for the beams with the long notch ($a = 0.5$) and four sizes $d = 50$~mm, $100$~mm, $200$~mm and $400$~mm. Error bars of the numerical results show the mean plus and minus one standard deviation of 100 analyses.}
\label{fig:lCmodLongNotch}
\end{figure}

\begin{figure}
\begin{center}
\includegraphics[width=10cm]{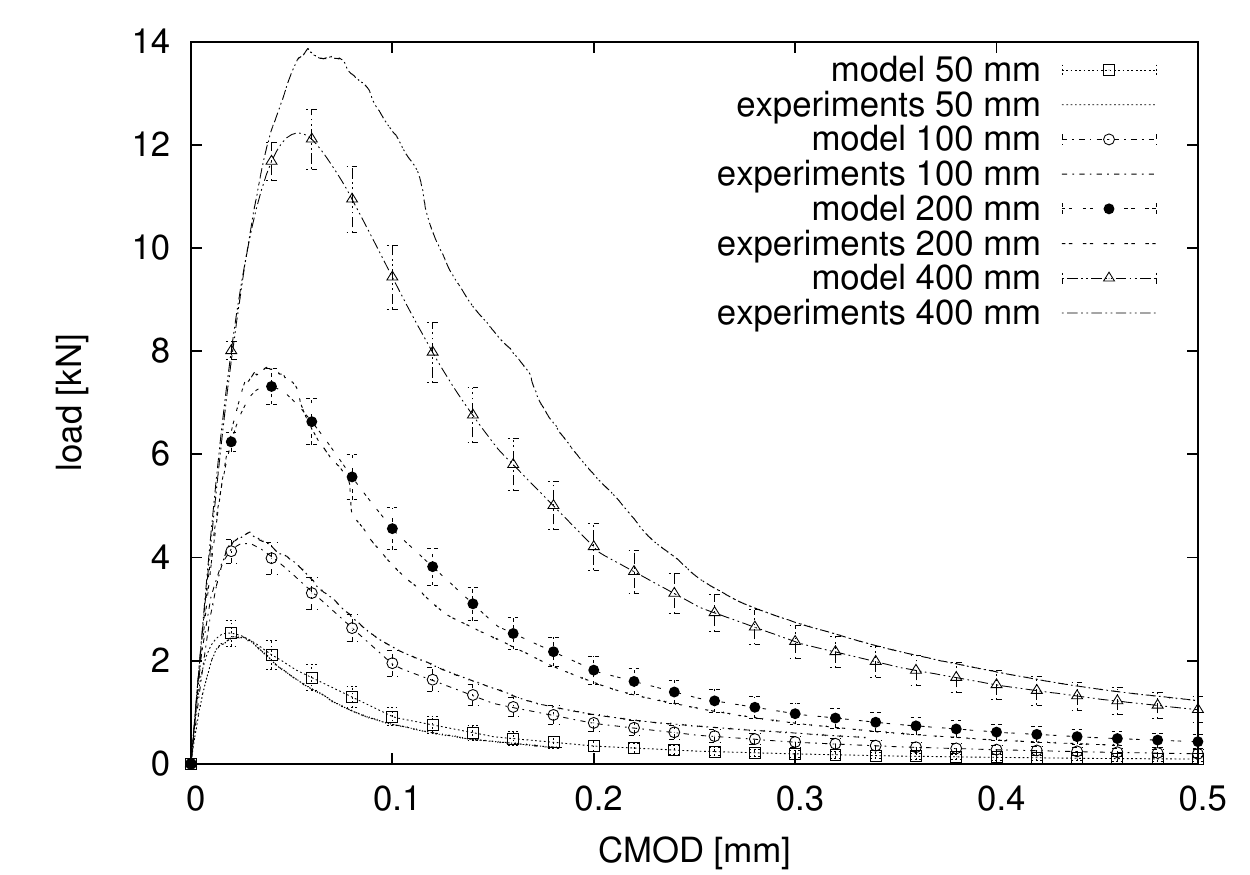}
\end{center}
\caption{Comparison of load versus CMOD of analyses and experiments for the beams with the short notch ($a = 0.2$) and four sizes $d = 50$~mm, $100$~mm, $200$~mm and $400$~mm. Error bars of the numerical results show the mean plus and minus one standard deviation of 100 analyses.}
\label{fig:lCmodShortNotch}
\end{figure}

\begin{figure}
\begin{center}
\includegraphics[width=10cm]{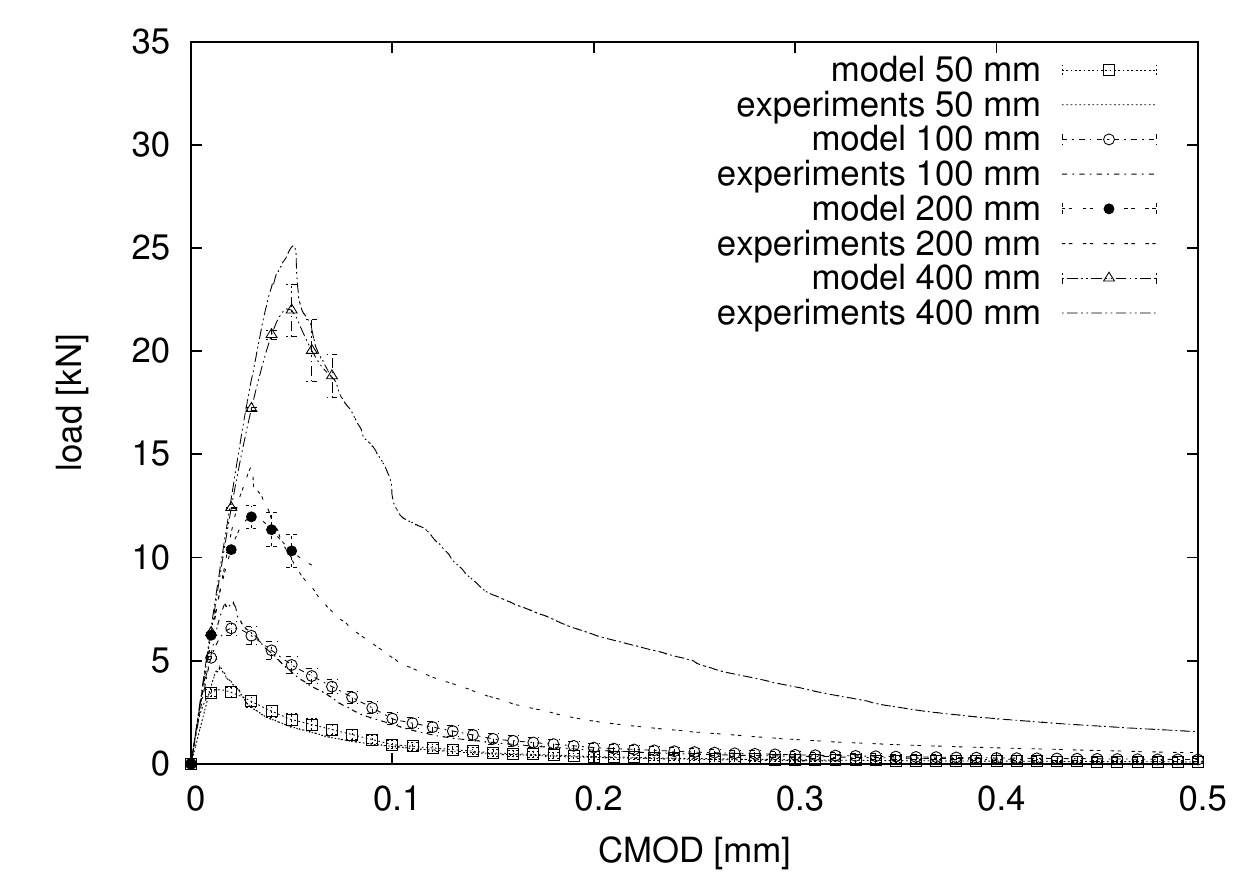}
\end{center}
\caption{Comparison of load versus CMOD of analyses and experiments for the beams without notch ($a = 0$) and four sizes $d = 50$~mm, $100$~mm, $200$~mm and $400$~mm. Error bars of the numerical results show the mean plus and minus one standard deviation of 100 analyses.}
\label{fig:lCmodUnnotched}
\end{figure}

The load-CMOD curves obtained in the analyses for long notched, short notched and unnotched beams for the four sizes are compared with experimental results in Figs.~\ref{fig:lCmodLongNotch},~\ref{fig:lCmodShortNotch}~and~\ref{fig:lCmodUnnotched}, respectively.
The load-CMOD curves presented are averages of 100 analyses with random meso-structures and autocorrelated fields.
The error bars indicate the mean plus and minus one standard deviation obtained in the meso-scale analyses.
For the experiments, standard deviations are not presented, since only a small number of beams for each size and geometry was tested, rendering any estimate of standard deviation inaccurate \citep{GreSolPij11}.
The agreement between numerical and experimental results for both the peak and post-peak part is very good considering that the same set of model parameters were used for obtaining all twelve load-CMOD curves.
In particular, the shape of the post-peak part of the load-CMOD curve is in very good agreement with the experimental results.  
However, some differences are present. 
For the long notched beams, the analyses overestimate the peaks observed in experiments for the small sizes (Fig.~\ref{fig:lCmodLongNotch}).
On the other hand, for the short notched and unnotched beams, the numerical analyses underestimate the peaks obtained in the experiments for the largest specimen (Figs.~\ref{fig:lCmodShortNotch}~and~\ref{fig:lCmodUnnotched}).
Overall, the model shows a tendency to overestimate the size effect observed in experiments. 

\section{Analysis of fracture processes}

In addition to the global results in the form of load-CMOD curves, local results such as damage patterns and spatial distributions of dissipated energy densities were studied. Firstly, the procedure to determine these results are illustrated for the largest specimen with the long notch ($a=0.5$, $d=400$~mm). 
Then selected distributions of dissipated energy densities are compared for different beam sizes and boundary conditions.

\begin{figure}
\begin{center}
\includegraphics[width=10cm]{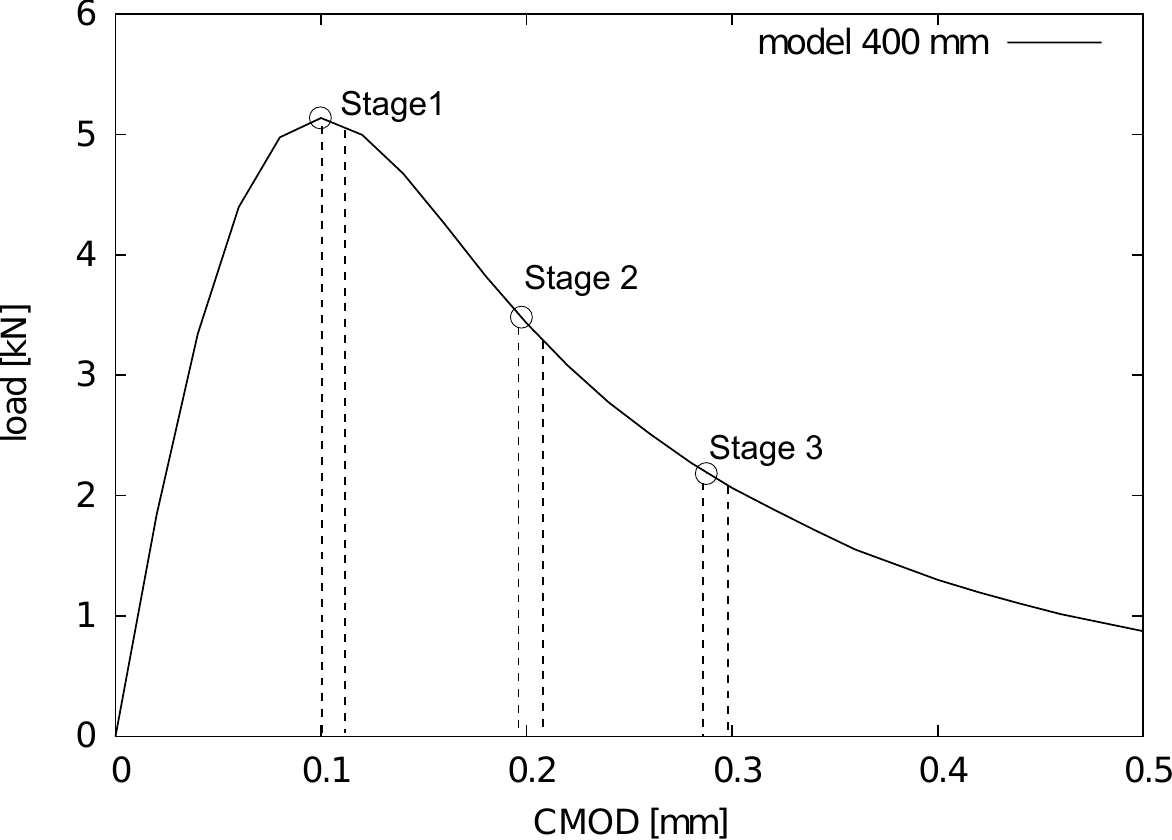}
\end{center}
\caption{Load-CMOD curve (average of 100 analyses) for the long notch and largest size ($a= 0.5$, $d = 400$~mm) with indications of three stages and CMOD increments used for the analyses of evolution of the fracture process zone.}
\label{fig:lDLN400Evol}
\end{figure}

\begin{figure}
\begin{center}
\begin{tabular}{ccc}
\includegraphics[height=6cm]{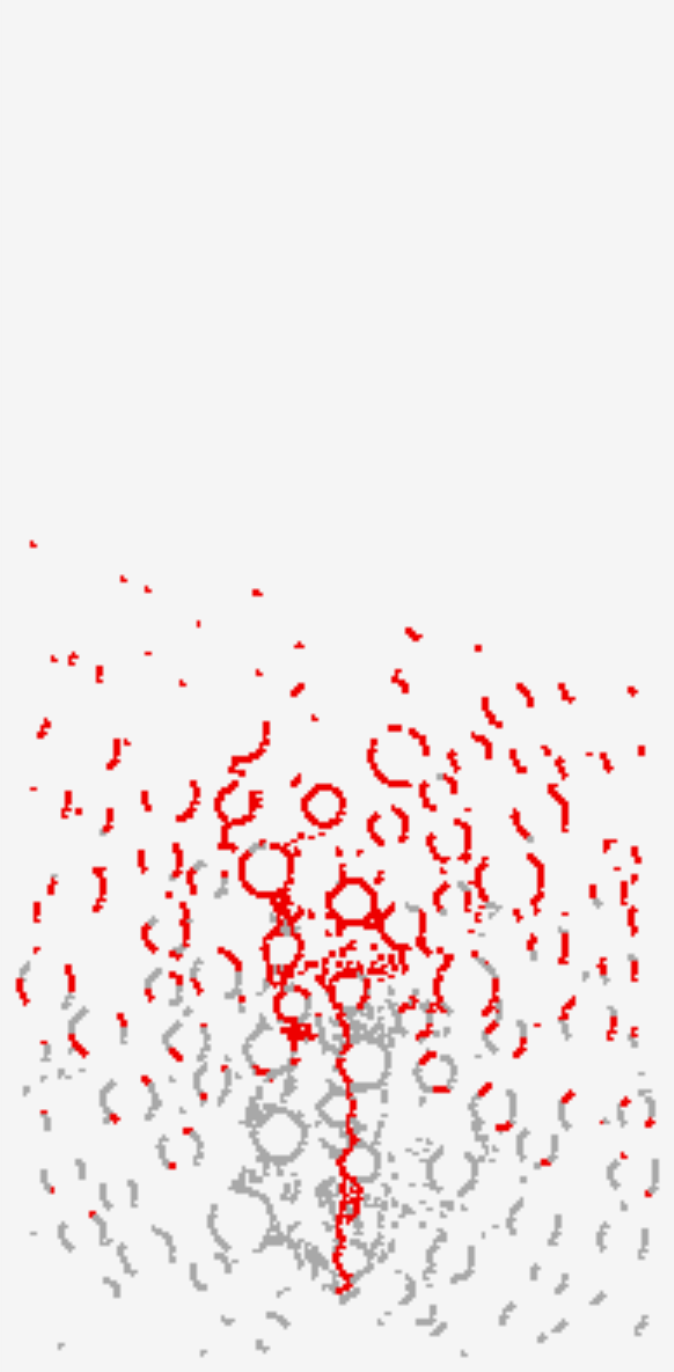} & \includegraphics[height=6cm]{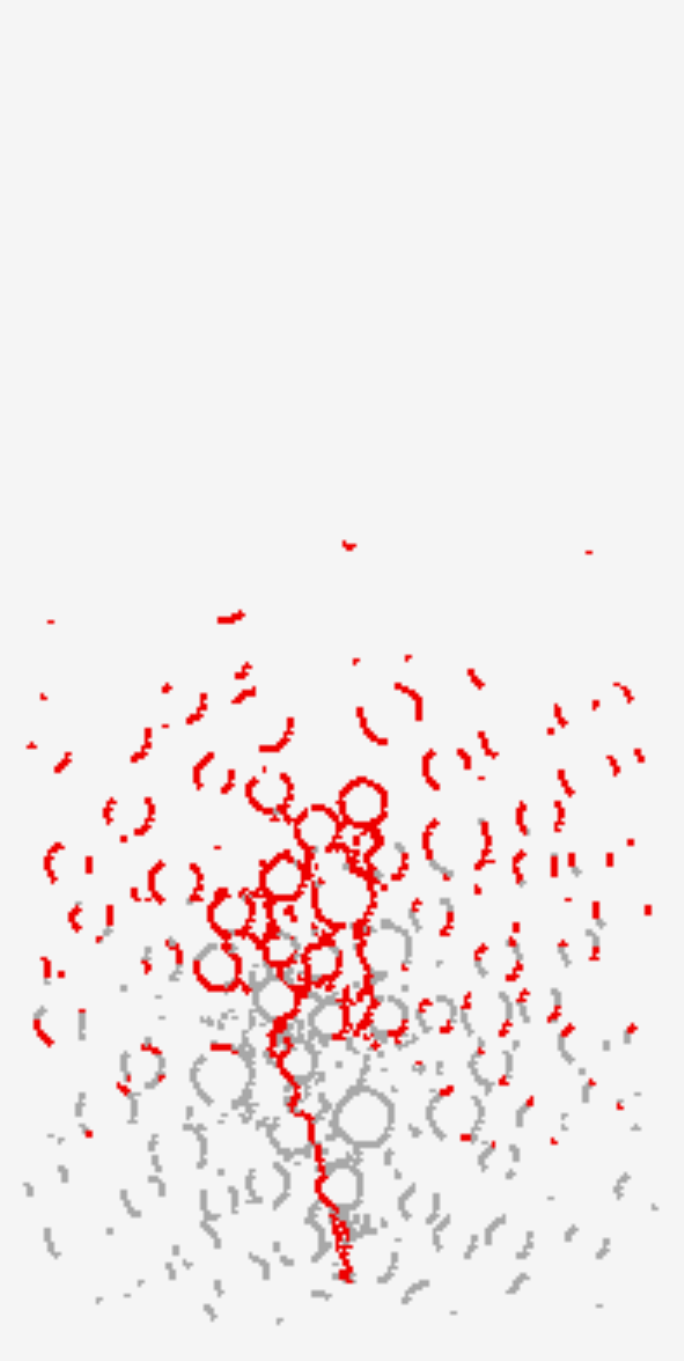} & \includegraphics[height=6cm]{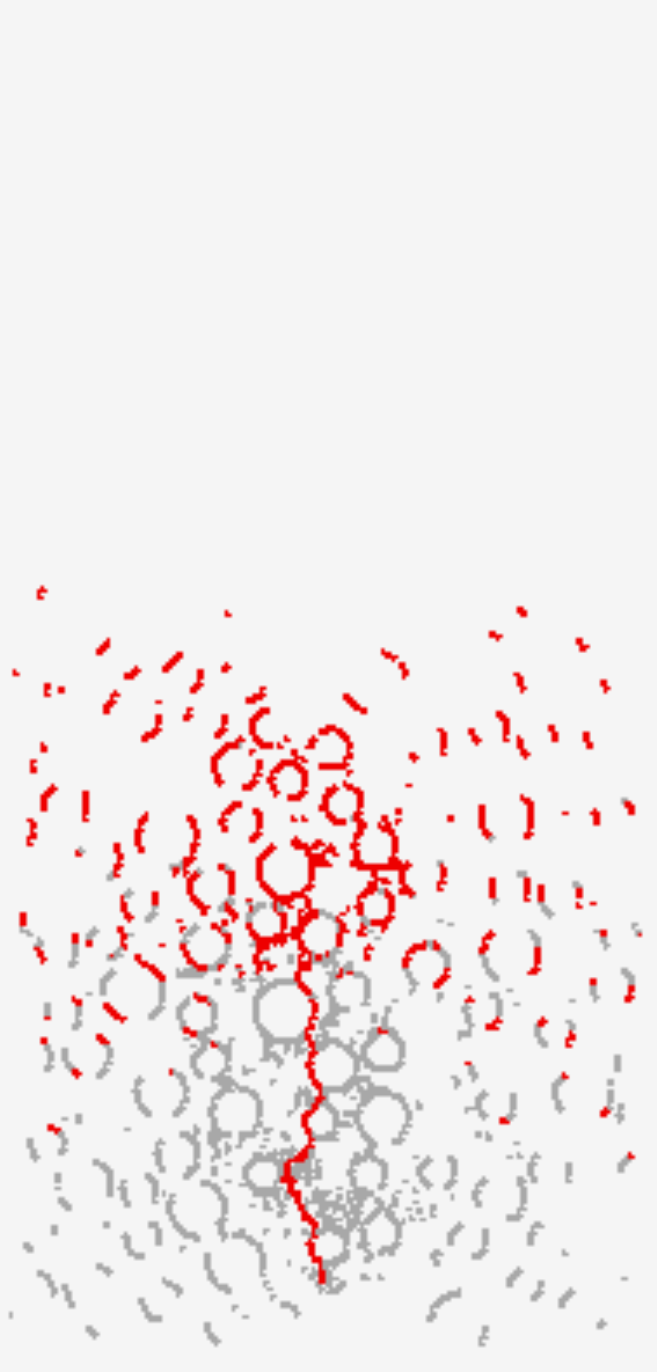}\\
(a) & (b) & (c)   
\end{tabular}
\end{center}
\caption{Damage patterns for the largest beam $d=400$~mm with the long notch ($a = 0.5$) for stage 1 (Fig.~\ref{fig:lDLN400Evol}) for three random analyses. Red (dark grey) lines indicate middle cross-sections with increasing damage at this stage. Light grey lines indicate middle cross-sections of damaged elements, in which damage does not increase at this stage.}
\label{fig:crackEvolStep1}
\end{figure}

\begin{figure}
\begin{center}
\begin{tabular}{ccc}
\includegraphics[height=6cm]{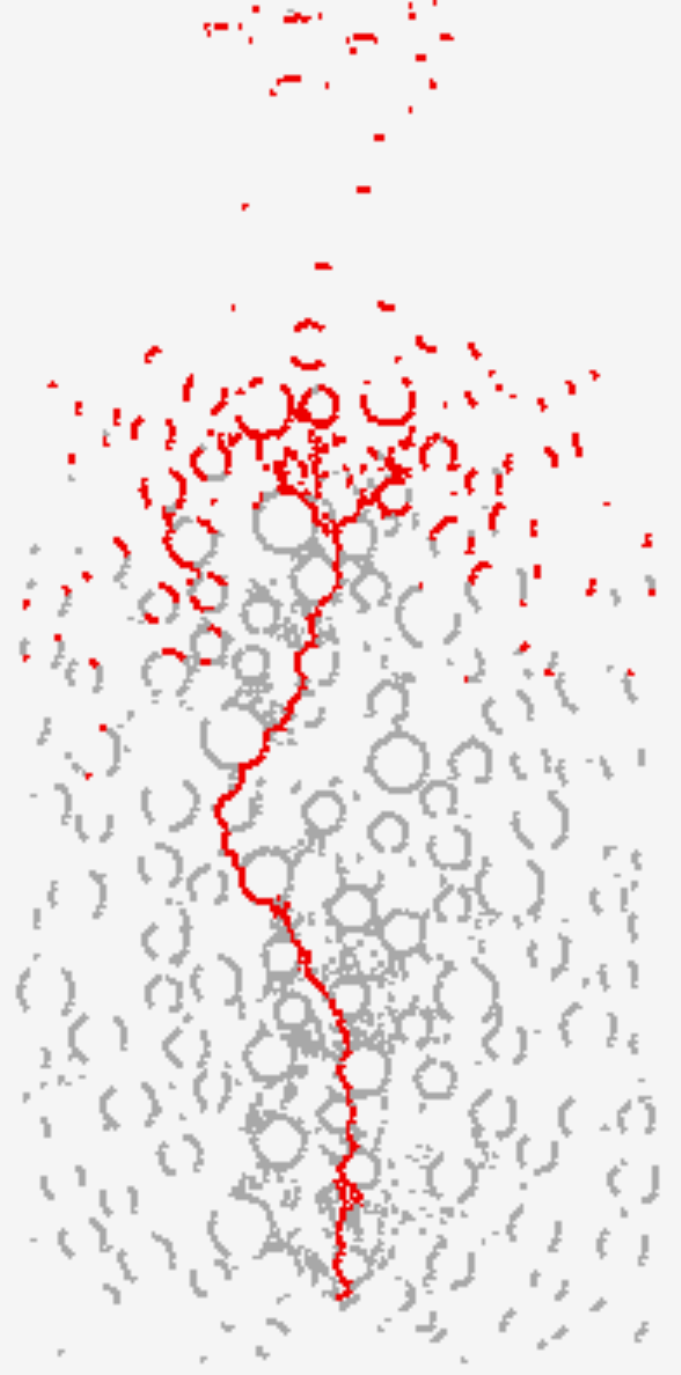} & \includegraphics[height=6cm]{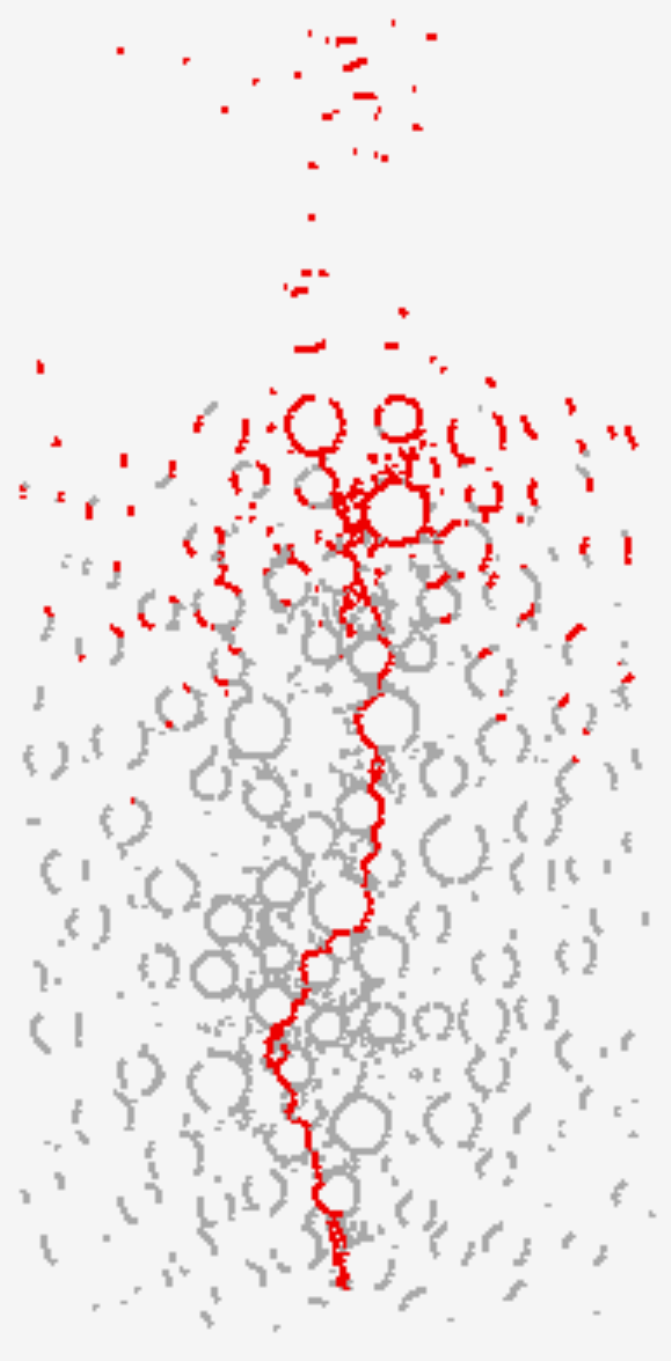} &  \includegraphics[height=6cm]{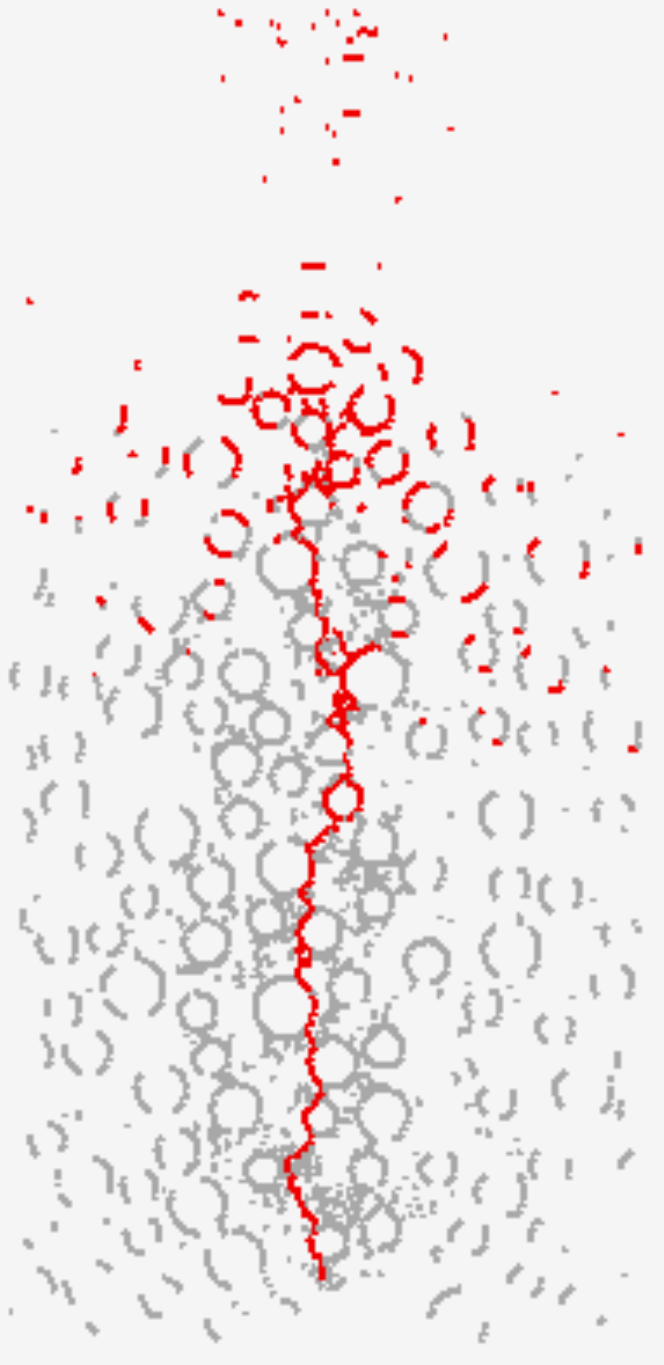}\\
(a) & (b) & (c)
\end{tabular}
\end{center}
\caption{Damage patterns for the largest beam $d=400$~mm with the long notch ($a = 0.5$) for stage 2 (Fig.~\ref{fig:lDLN400Evol}) for three random analyses. Red (dark grey) lines indicate middle cross-sections with increasing damage at this stage. Light grey lines indicate middle cross-sections of damaged elements, in which damage does not increase at this stage.}
\label{fig:crackEvolStep2}
\end{figure}

\begin{figure}
\begin{center}
\begin{tabular}{ccc}
\includegraphics[height=6cm]{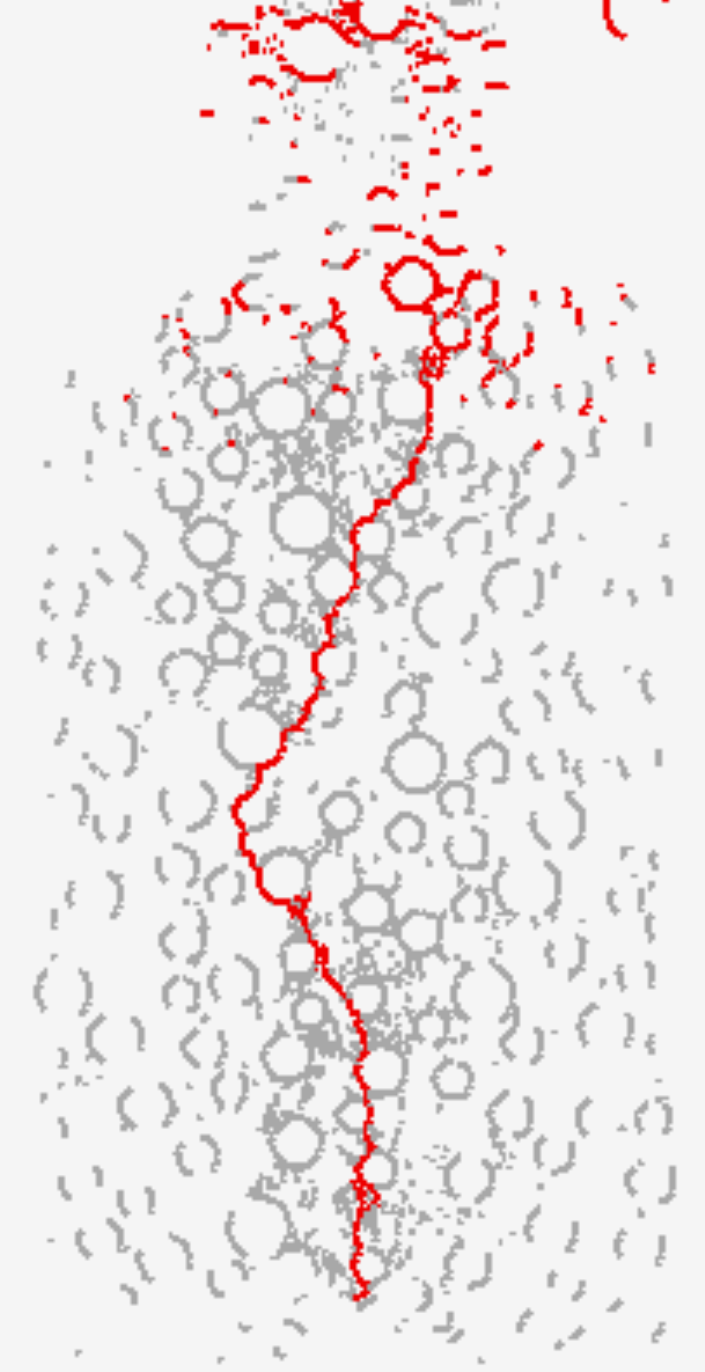} &  \includegraphics[height=6cm]{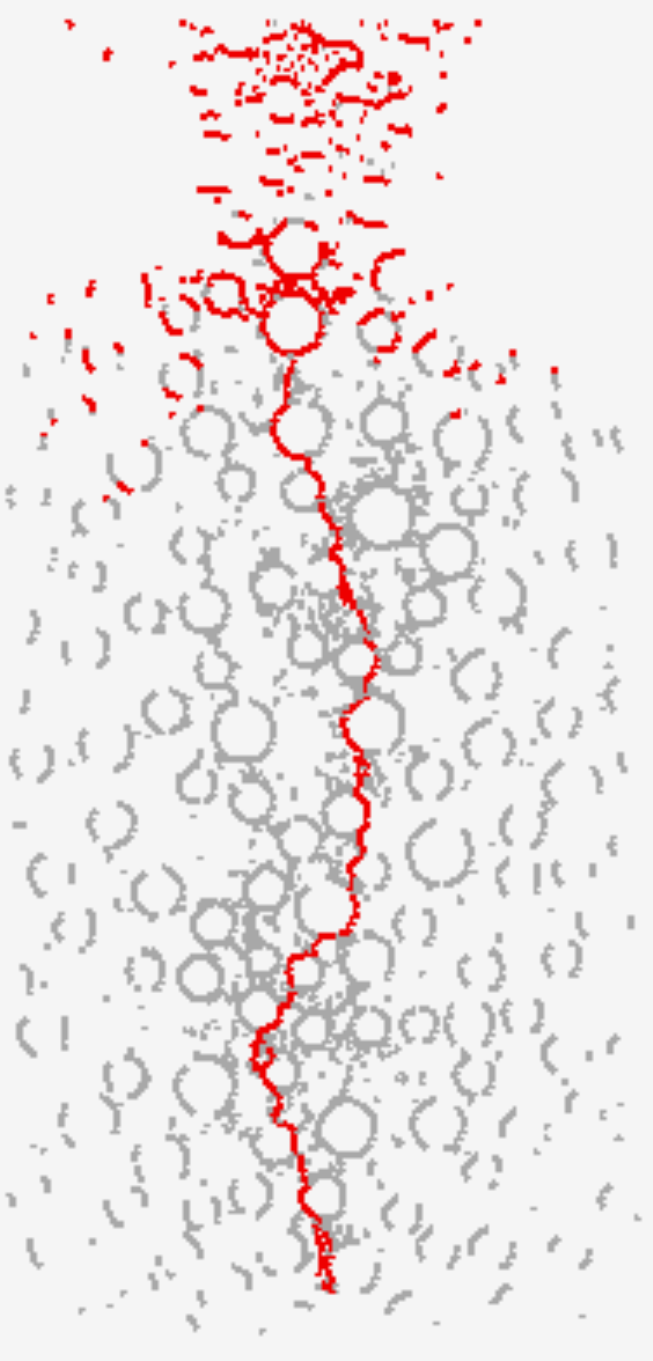} & \includegraphics[height=6cm]{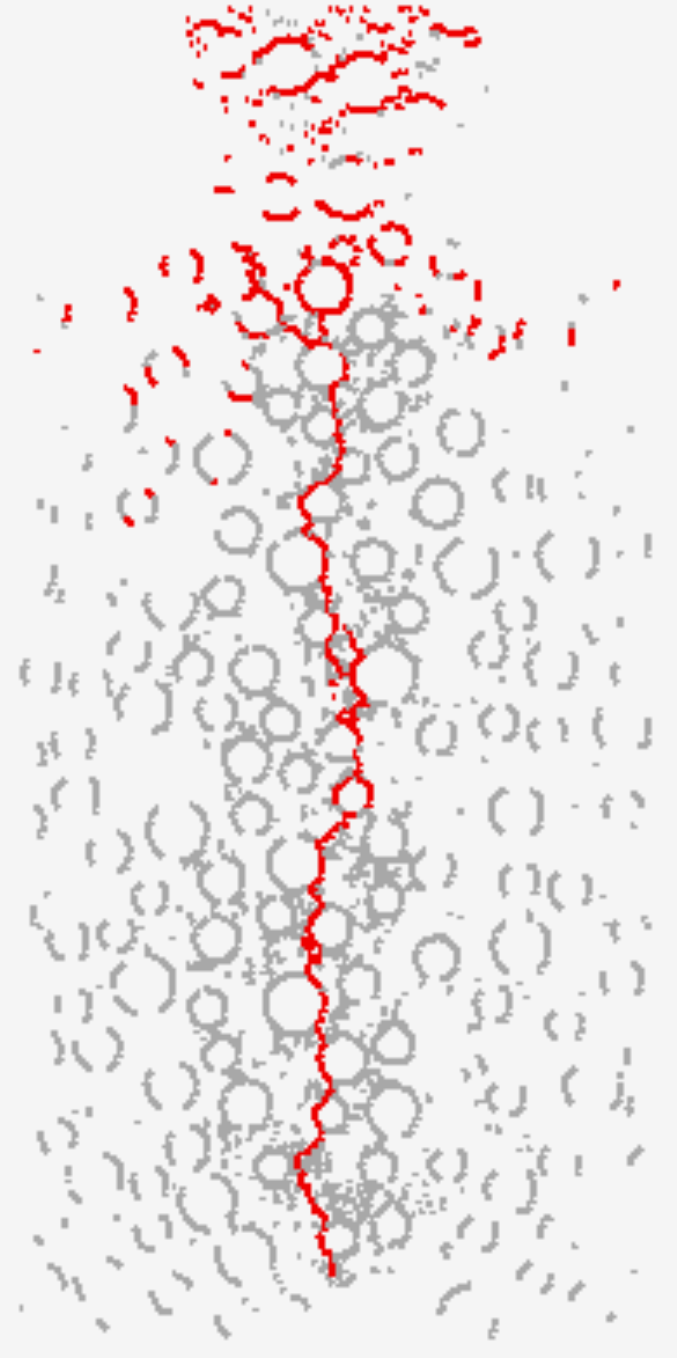}\\
(a) & (b) & (c)
\end{tabular}
\end{center}
\caption{Damage patterns for the largest beam $d=400$~mm with the long notch ($a = 0.5$) for stage 3 (Fig.~\ref{fig:lDLN400Evol}) for three random analyses. Red (dark grey) lines indicate middle cross-sections with increasing damage at this stage. Light grey lines indicate middle cross-sections of damaged elements, in which damage does not increase at this stage.}
\label{fig:crackEvolStep3}
\end{figure}

\begin{figure}
\begin{center}
\includegraphics[width=10cm]{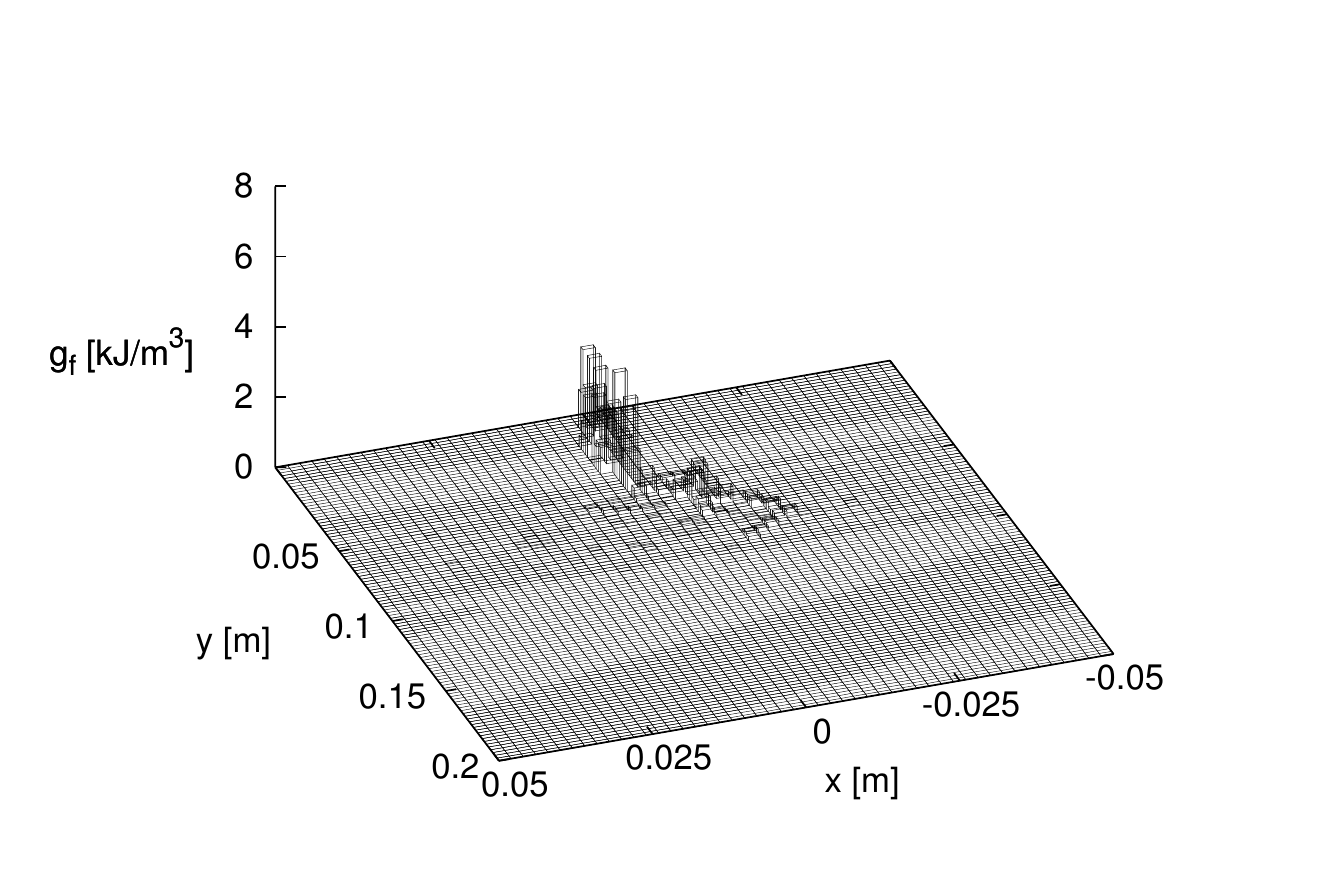}\\
(a)\\
\includegraphics[width=10cm]{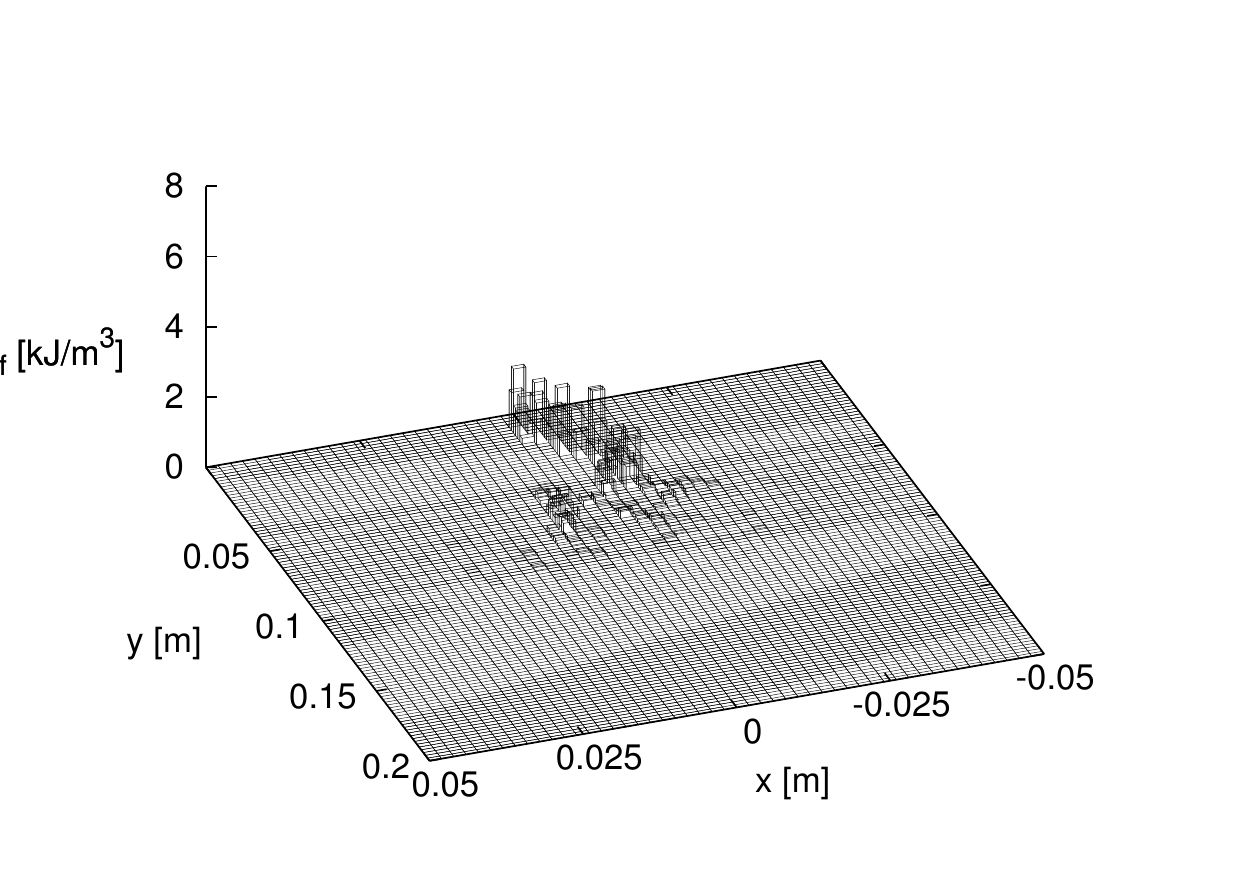}\\
(b)\\
\includegraphics[width=10cm]{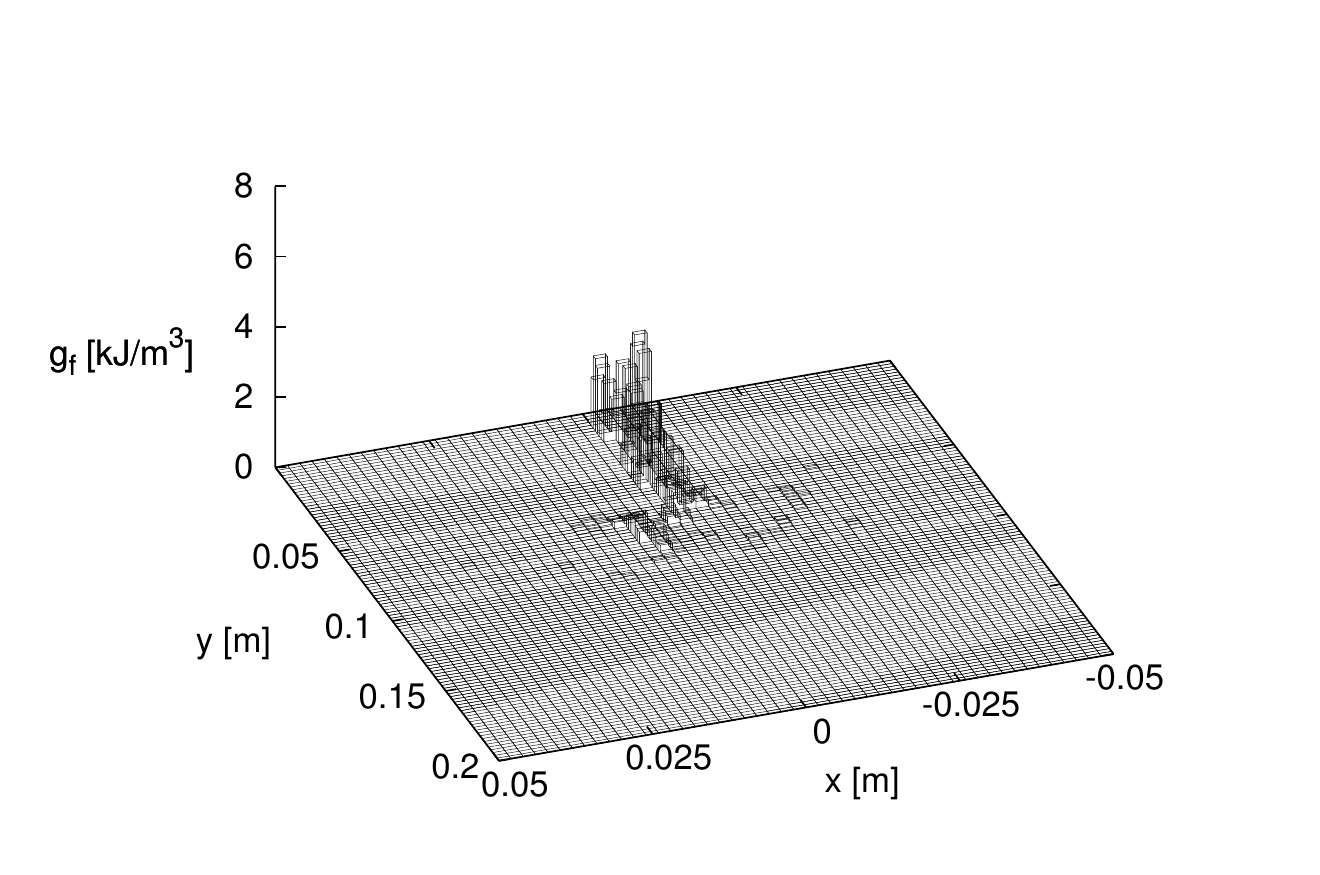}\\
(c)\\
\end{center}
\caption{Dissipated energy densities for the same 3 random analyses as in Fig.\ref{fig:crackEvolStep1} for the largest beam $d=400$~mm with the long notch ($a = 0.5$) for CMOD increment 1 (Fig.~\ref{fig:lDLN400Evol}).}
\label{fig:dissChange1}
\end{figure}

\begin{figure}
\begin{center}
\includegraphics[width=10cm]{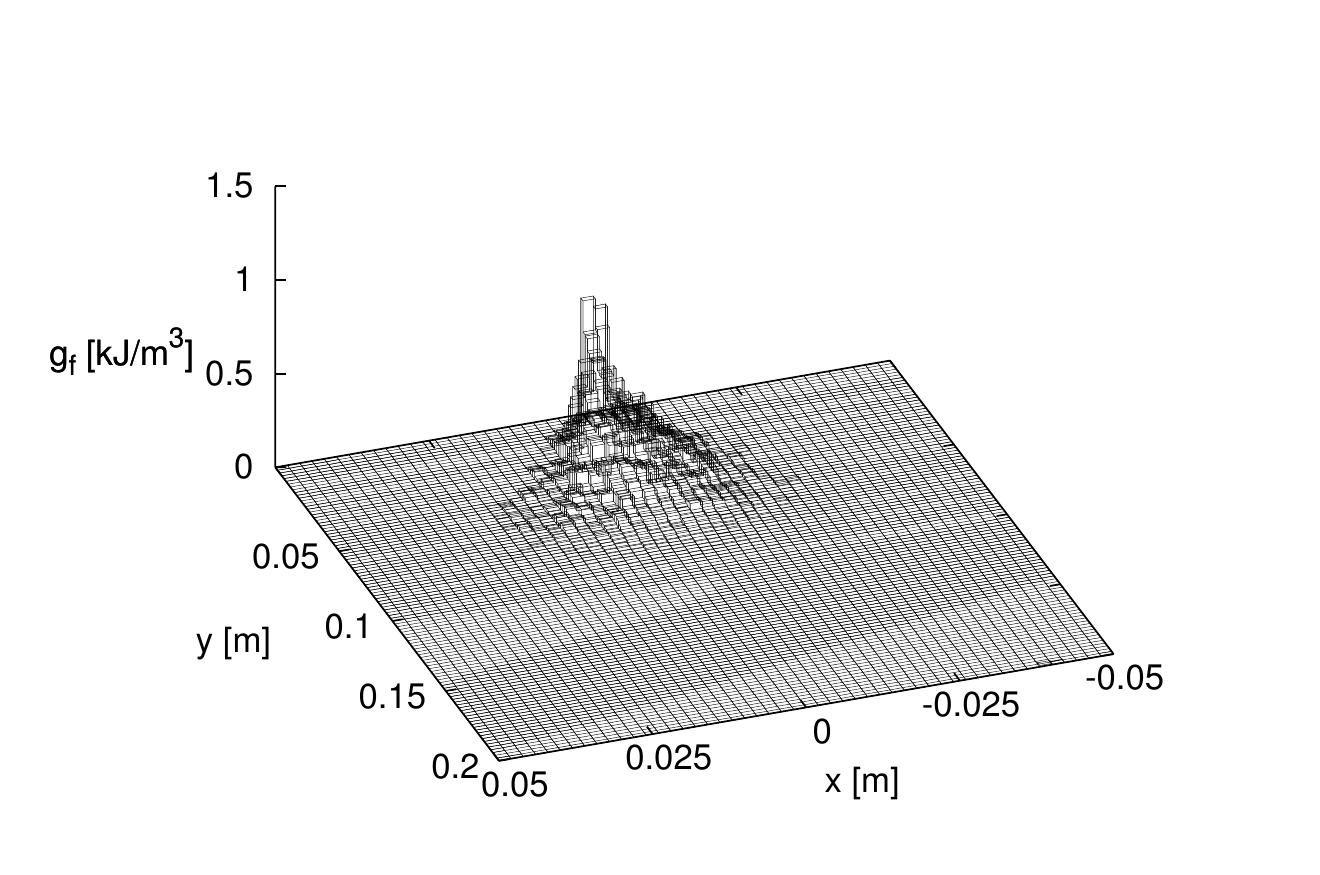}\\
(a)\\
\includegraphics[width=10cm]{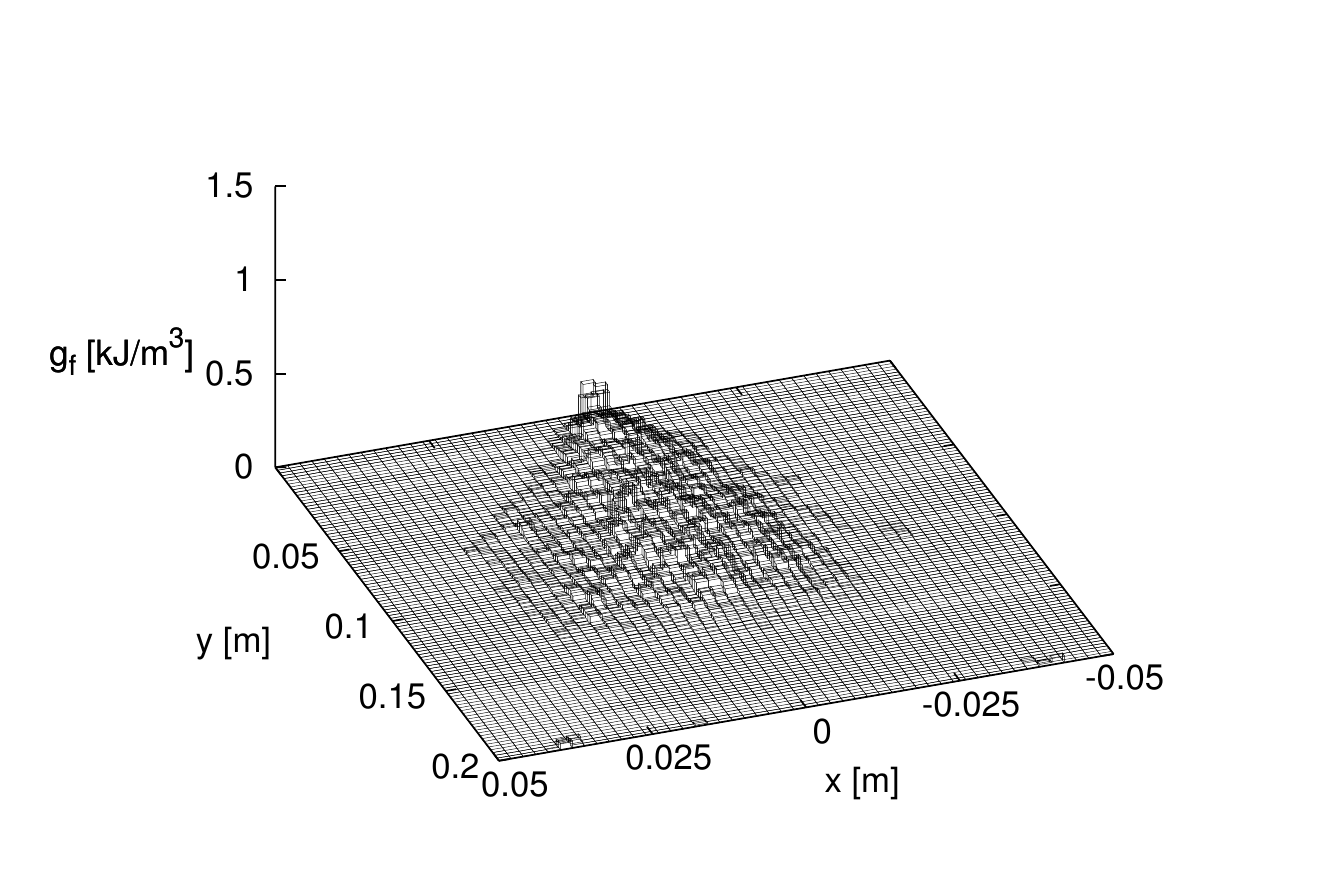}\\
(b)\\
\includegraphics[width=10cm]{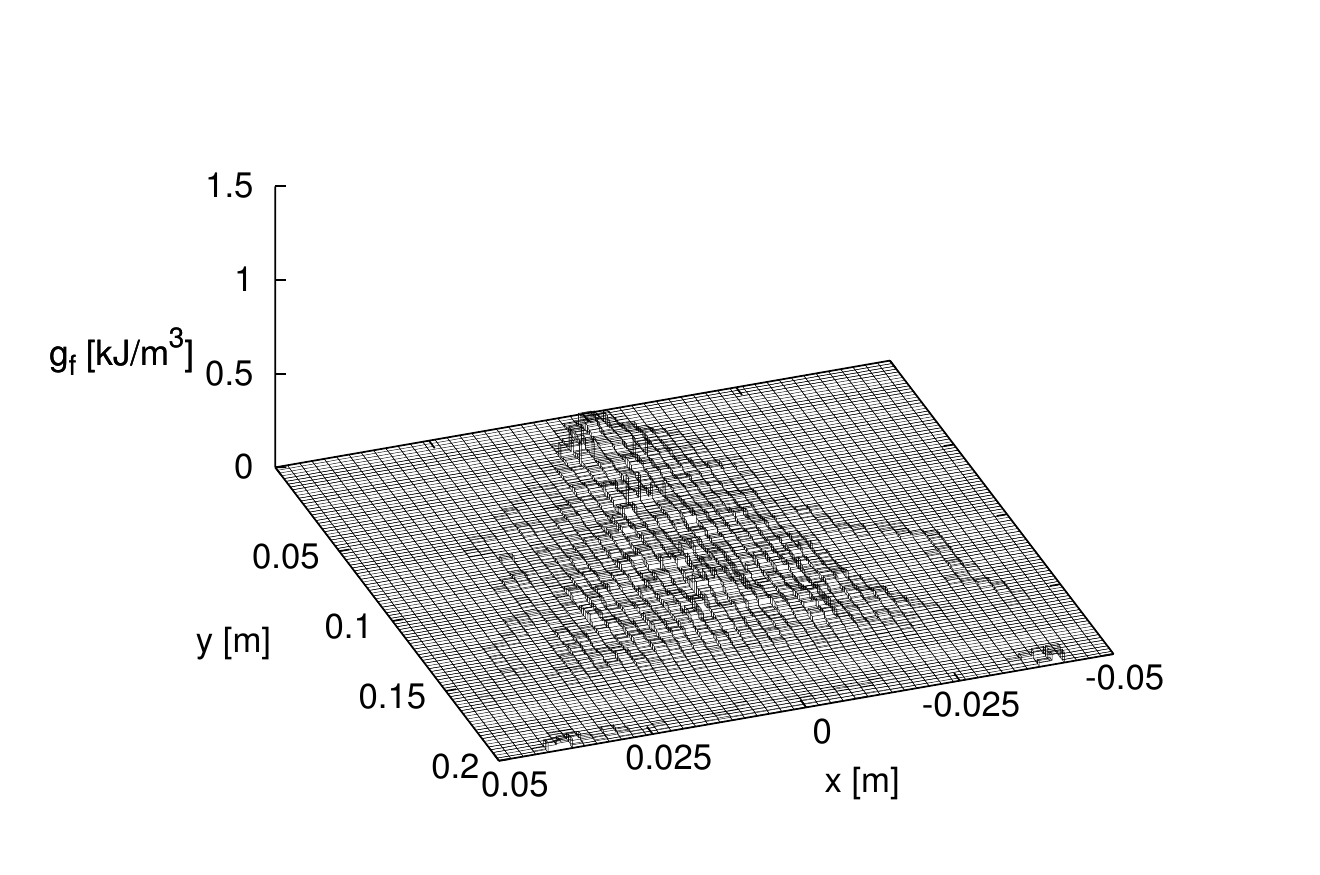}\\
(c)\\
\end{center}
\caption{a) Mean of dissipated energy densities of 100 random analyses for the largest beam $d=400$~mm with the long notch ($a = 0.5$) for (a) CMOD increment 1, (b) CMOD increment 2 and (c) CMOD increment 3 (Fig.~\ref{fig:lDLN400Evol}).}
\label{fig:dissChange2}
\end{figure}

In the nonlinear analyses, the fracture process is modelled by damage in the lattice elements. The evolution of the resulting damage patterns was studied for the long notch beam for three stages marked in the mean load-CMOD curve shown in Fig.~\ref{fig:lDLN400Evol}. 
The damage patterns are shown in Figs.~\ref{fig:crackEvolStep1}~to~\ref{fig:crackEvolStep3}, respectively, for three random analyses for stages~1,~2~and~3. 
Red (dark grey) lines mark the middle cross-sections of elements in which damage increases at this stage of the analysis. On the other hand, grey lines indicate middle cross-sections of elements in which damage increased at an earlier stage of analyses, but did not increase at this stage. 
From the damage patterns it can be seen that the fracture process zone obtained from the analyses consists of a localised tortuous row of cross-sections of elements, in which damage increases. 
The location of this row of damaged elements differs considerably for the three random analyses.
Therefore, it is required to average the response of the random analyses to be able to compare the results of beams of different sizes and geometries. 
Otherwise, it would not be possible to distinguish between dependencies originating from the random meso-structure and specimen size and boundary conditions.
Dissipated energy density, which is a physical quantity that can be integrated and averaged, was used to analyse the results following the approach proposed in \citet{GraJir10}.
The increment of dissipated energy in a lattice element of length $h$ was calculated as 
\begin{equation}
\Delta D_{\rm d} = \Delta \omega A h \dfrac{1}{2} \boldsymbol{\varepsilon} \mathbf{D} \boldsymbol{\varepsilon}
\end{equation}
where $\Delta \omega$ is the increment of the damage parameter.
This expression is a good approximation of the rate of dissipated energy, if the increment of damage is very small. Therefore, subincrementation was applied for the evaluation of the dissipated energy, so that the accuracy of the computed value is independent of the CMOD increments used to control the analysis.
For the averaging of dissipated energy, the domain to be analysed was discretised by a square grid, with a cell size of $3.125$~mm, which is greater than $d_{\rm min} = 1$~mm used to determine the length of lattice elements.
For this grid, the locations of the mid cross-section of lattice elements determine the amount of energy that is dissipated in the cells.
If a mid cross-section of an element is located in multiple cells, the energy is allocated in proportion to the section of the mid cross-section that is located in each cell.
In the next step, the energy density of a cell is determined as the sum of energy divided by the cell area.
The spatial distributions of dissipated energy densities for the increment of CMOD just after stage 1 (Fig.~\ref{fig:lDLN400Evol}) for the same three random analysis presented before is shown in Figs.~\ref{fig:dissChange1}a,~b~and~\ref{fig:dissChange2}a.
This spatial distribution of dissipated energy density represents the fracture process zone at this stage of analysis.
The majority of energy is dissipated in a localised region, which is tortuous. This has already been observed by studying the damage patterns in Figs.~\ref{fig:crackEvolStep1}~to~\ref{fig:crackEvolStep3}. Similar results have been reported in the literature \citep{CedDeiIor87,PlaEli93b,NirHor92,BolHikShi93}.
To be able to compare the distribution of dissipated energy densities for different sizes and geometries, the results of 100 random analyses were averaged for an increment of CMOD at stage 1 (Fig.~\ref{fig:dissChange2}a). 
This average dissipated energy density is distributed over a wider zone compared to the densities of the individual random zones, whereby the width of the average distribution is mainly determined by the tortuosity of the fracture process zone of individual analyses. 
The maximum value of the density is at the notch ($x=y=0$). See Fig.~\ref{fig:geometry} for the definition of the coordinate system. 
Further along the ligament ($y>0$) the fracture process zone widens and the density decreases.
In Figs.~\ref{fig:dissChange2}b~and~c, the average density distributions for CMOD increments at stage 2 and 3 are shown. All these CMOD increments were chosen so that the mean dissipated energy increment is equal to approximately 0.0115 J. 
The values of dissipated energy is much higher in Fig.~\ref{fig:dissChange1} than in Fig.~\ref{fig:dissChange2}, since for the individual analyses the zone in which energy is dissipated is much smaller than the corresponding zone for the average distribution. 

\begin{figure}
\begin{center}
\includegraphics[width=10cm]{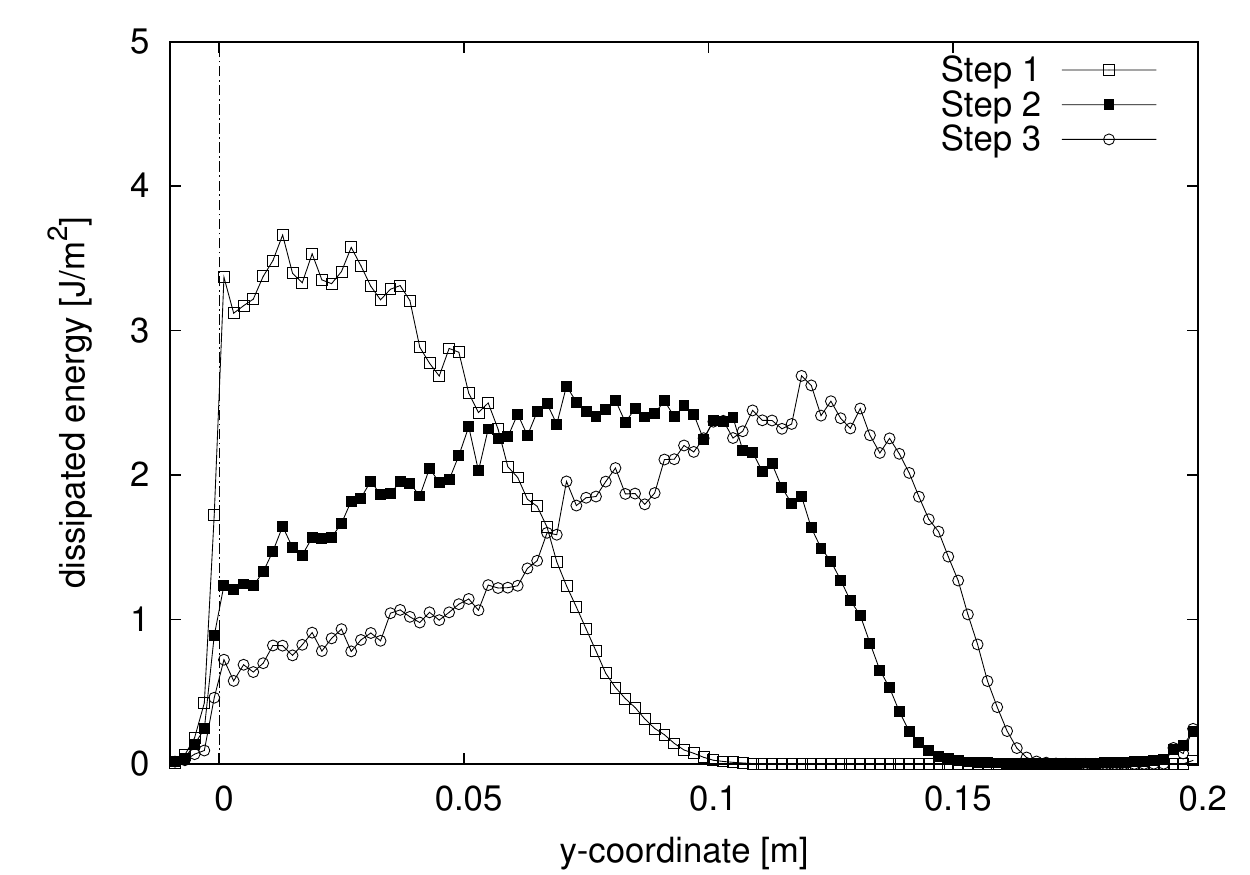}
\end{center}
\caption{Evolution of the dissipated energy density along the ligament (y-direction) for the largest beam ($d=400$~mm) with the long notch ($a = 0.5$) for three CMOD increments (Fig.~\ref{fig:lDLN400Evol}). The results are based on an average of 100 random analyses.}
\label{fig:rCurveEvolLN}
\end{figure}

\begin{figure}
\begin{center}
\includegraphics[width=10cm]{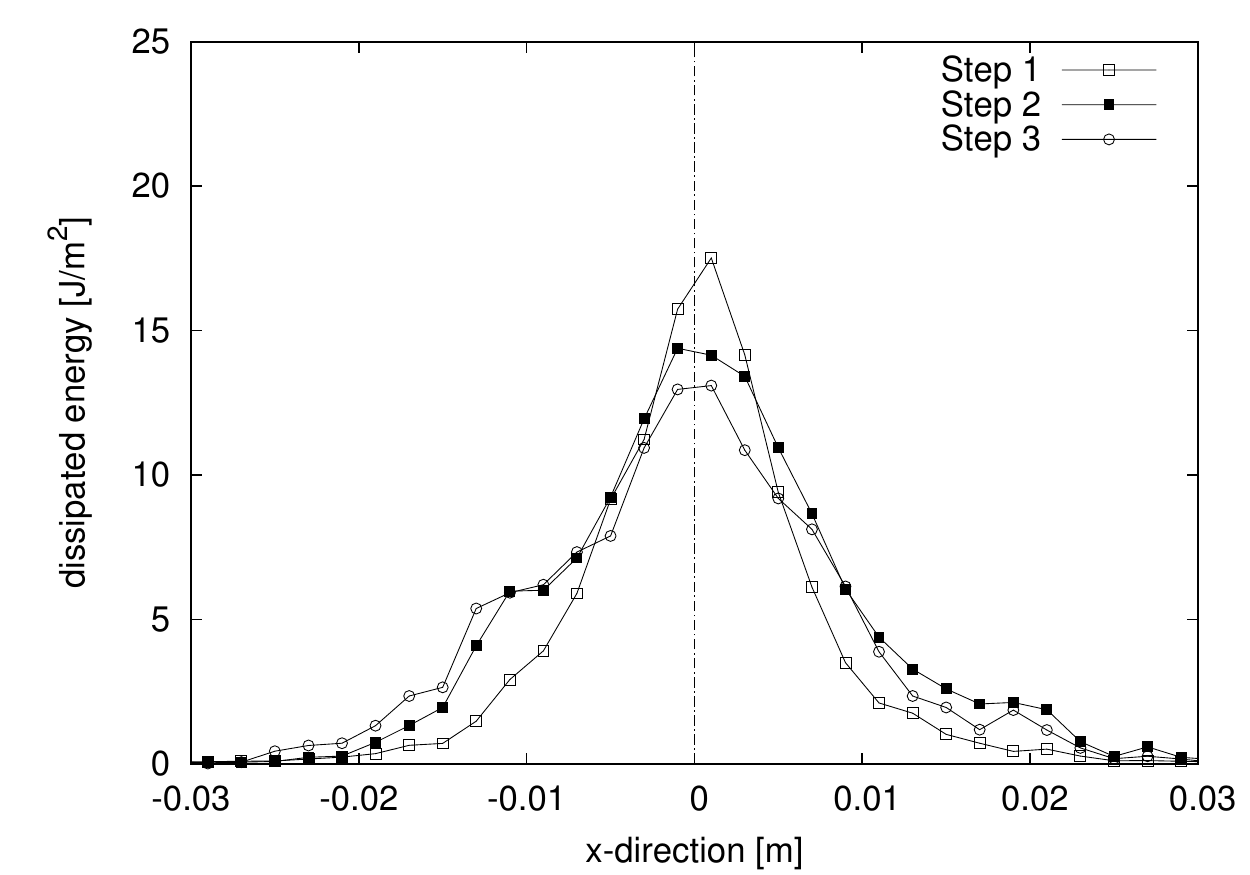}
\end{center}
\caption{Evolution of the dissipated energy density perpendicular to the ligament (x-direction) for the largest beam ($d=400$~mm) with the long notch ($a = 0.5$) for three CMOD increments indicated in Fig.~\ref{fig:lDLN400Evol}. The results are based on an average of 100 random analyses.}
\label{fig:rCurveOppEvolLN}
\end{figure}

These plots of energy density are suitable to illustrate the fracture process zone for one specimen and stage at a time.
However, for comparing quantitatively the results for different stages and geometries, it is suitable to reduce the dimensions of the plots.
Therefore, the fracture process zones are presented by the distribution of dissipated energy in the x and y-direction only.
These one-dimensional distributions are obtained by integrating the dissipated energy density in the opposite direction.
In Figs.~\ref{fig:rCurveEvolLN}~and~\ref{fig:rCurveOppEvolLN}, the two types of distribution of dissipated energy density are shown for the three increments marked in Fig.~\ref{fig:lDLN400Evol}. 

The fracture process zone travels along the ligament towards the top of the beam, whereas the distribution across the ligament remains almost constant at a width of around 4~cm.
At peak, the fracture process zone in the y-direction is characterised by high values of energy close to the notch ($0$~cm$<y<3$~cm) which decrease with increasing y-coordinate.
The length of the fracture process zone (y-direction) increases from around 9~cm at peak (stage~1) to around 17~cm in the post-peak regime (increment~3), where still significant dissipation close to the notch takes place. However, at increment 3 the density is greatest at around 13~cm, which is close to the tip of the fracture process zone.
Even at this stage, the present meso-scale model predicts energy dissipation close to the notch.

\begin{figure}
\begin{center}
\includegraphics[width=10cm]{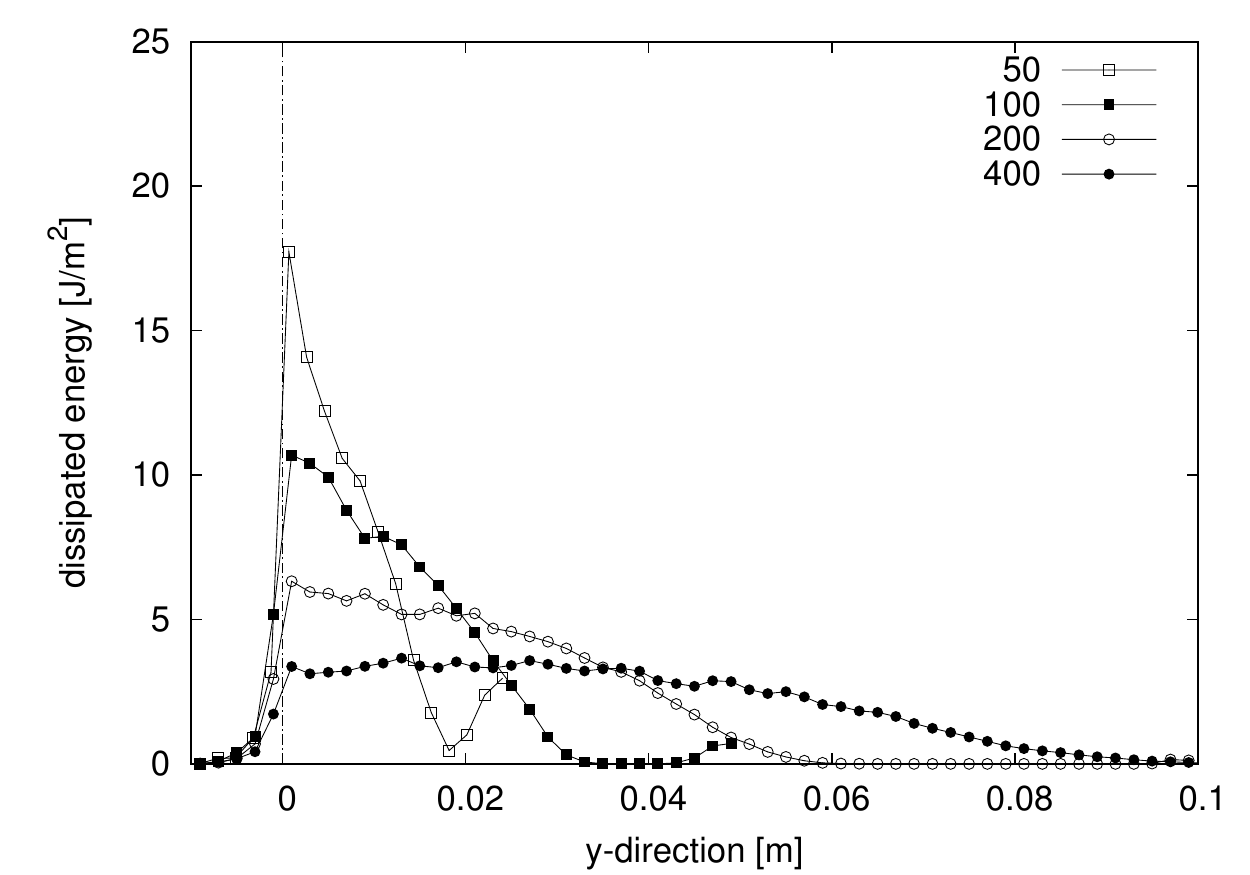}
\end{center}
\caption{Dissipated energy density increment along the ligament (y-direction) for the beams with long notch ($a = 0.5$) and four sizes $d = 50$, $100$, $200$ and $400$~mm for a CMOD increment just after the peak of the mean load-CMOD curves.  The results are based on an average of 100 random analyses.}
\label{fig:rCurveChangeCompLN}
\end{figure}

\begin{figure}
\begin{center}
\includegraphics[width=10cm]{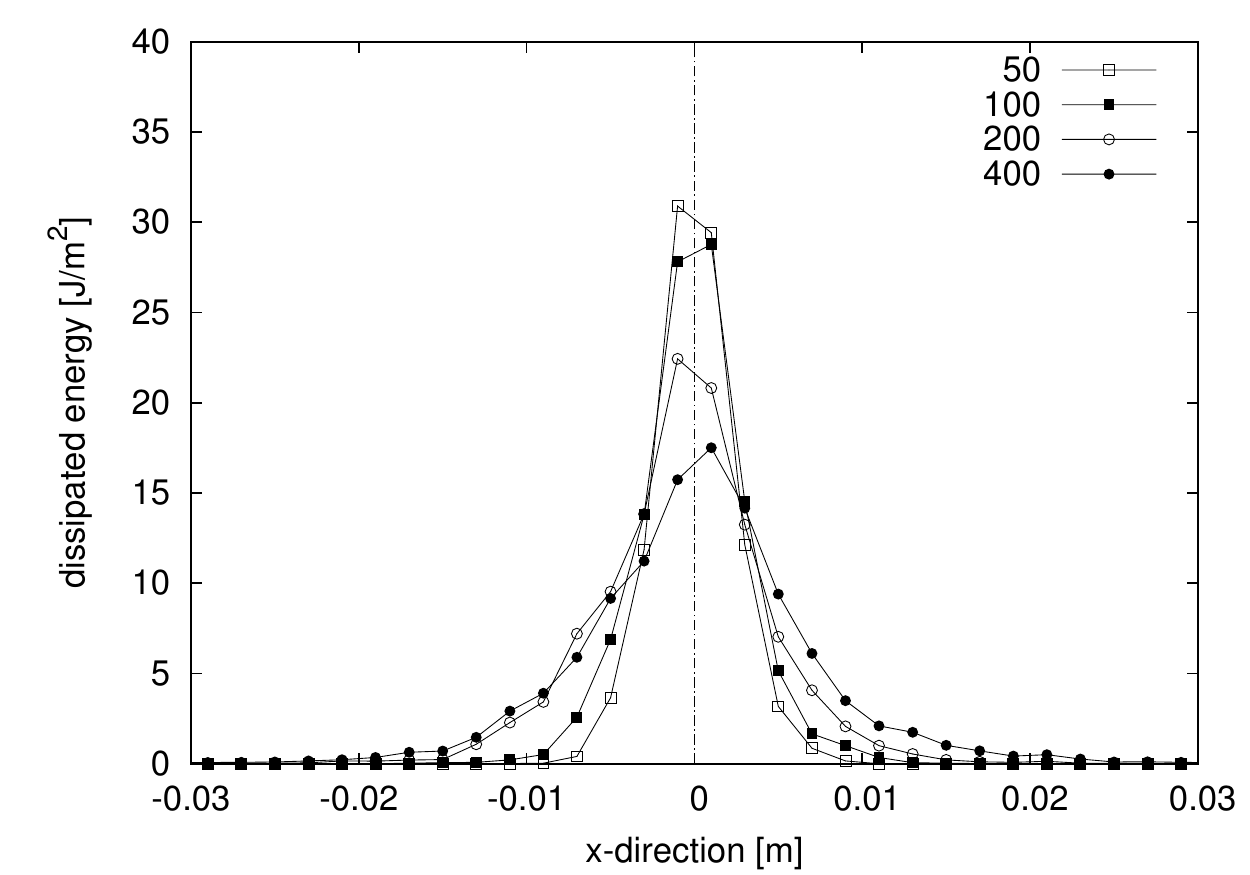}
\end{center}
\caption{Dissipated energy density increment perpendicular to the ligament (x-direction) for the long notch ($a = 0.5$) and four sizes $d = 50$, $100$, $200$ and $400$~mm for a CMOD increment just after the peak of the mean load-CMOD curves. The results are based on an average of 100 random analyses.}
\label{fig:rCurveChangeOppCompLN}
\end{figure}

So far, only results for one beam size and geometry have been presented.
To investigate how beam size and geometry influence the fracture process zones, the results for all other beams were evaluated as well.
The comparison of these results is limited to an increment of CMOD just after peak of the respective load-CMOD curves, which was chosen so that for all beams the corresponding mean dissipated energy increment is $\Delta D = 0.23$~J/m.
Firstly, the distributions for the four sizes of beams with the long notch are presented in Figs.~\ref{fig:rCurveChangeCompLN}~and~\ref{fig:rCurveChangeOppCompLN}.

The length of the fracture process zone at peak depends strongly on the size of the beam. The length increases from around 19~mm for the smallest beam (d=50~mm) to 82~mm for the greatest beam (d=400~mm).
This strong dependence of the length on the size can be explained by the decrease of the stress gradient with increasing beam size.
The width of the fracture process zone is much less dependent on the size of the beam.
It increases only slightly with increasing beam size. This slight increase originates from the increase of tortuosity with increasing length of fracture process zone for greater beams.

\begin{figure}
\begin{center}
\includegraphics[width=10cm]{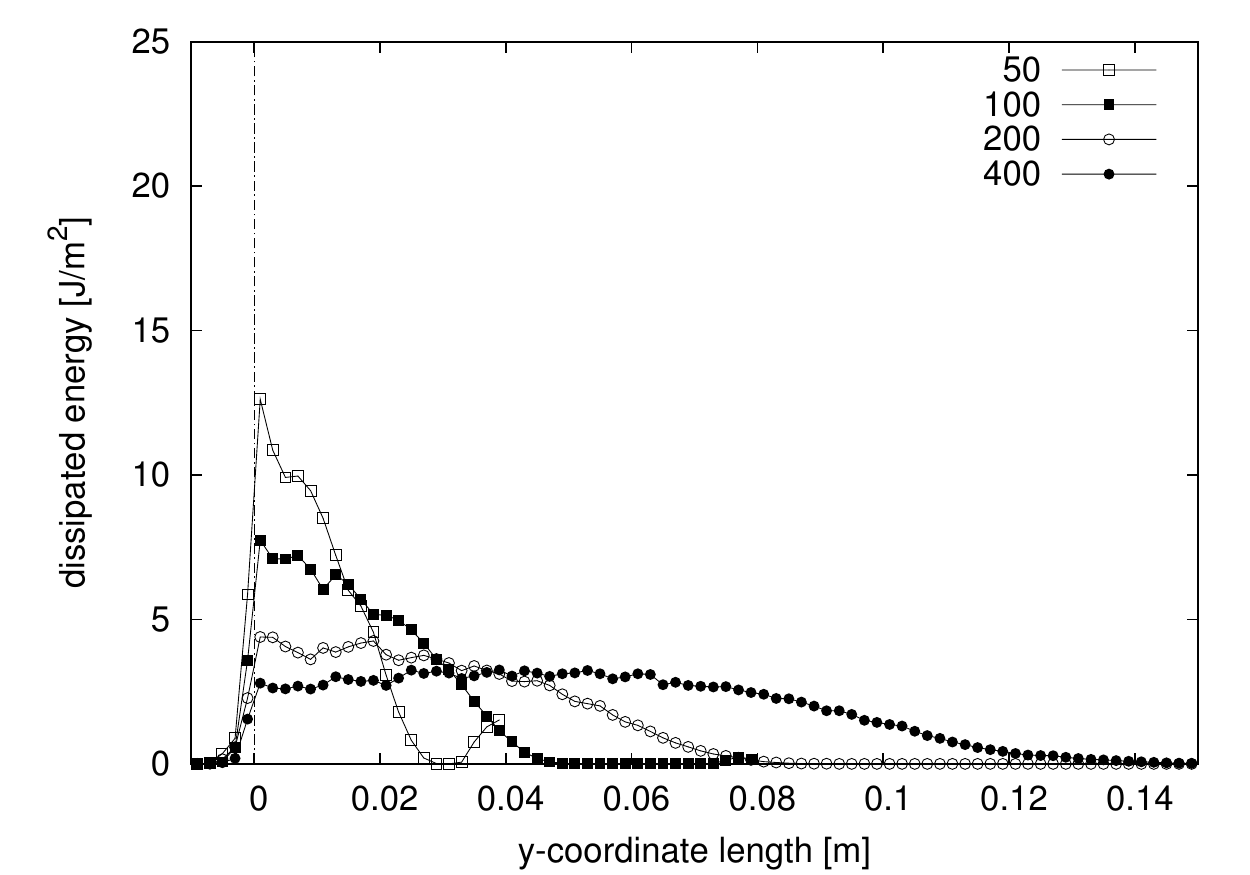}
\end{center}
\caption{Dissipated energy density increment along the ligament (y-direction) for the beams with short notch ($a = 0.2$) and four sizes $d = 50$, $100$, $200$ and $400$~mm for a CMOD increment just after the peak of the mean load-CMOD curves. The results are based on an average of 100 random analyses.}
\label{fig:rCurveChangeCompSN}
\end{figure}

\begin{figure}
\begin{center}
\includegraphics[width=10cm]{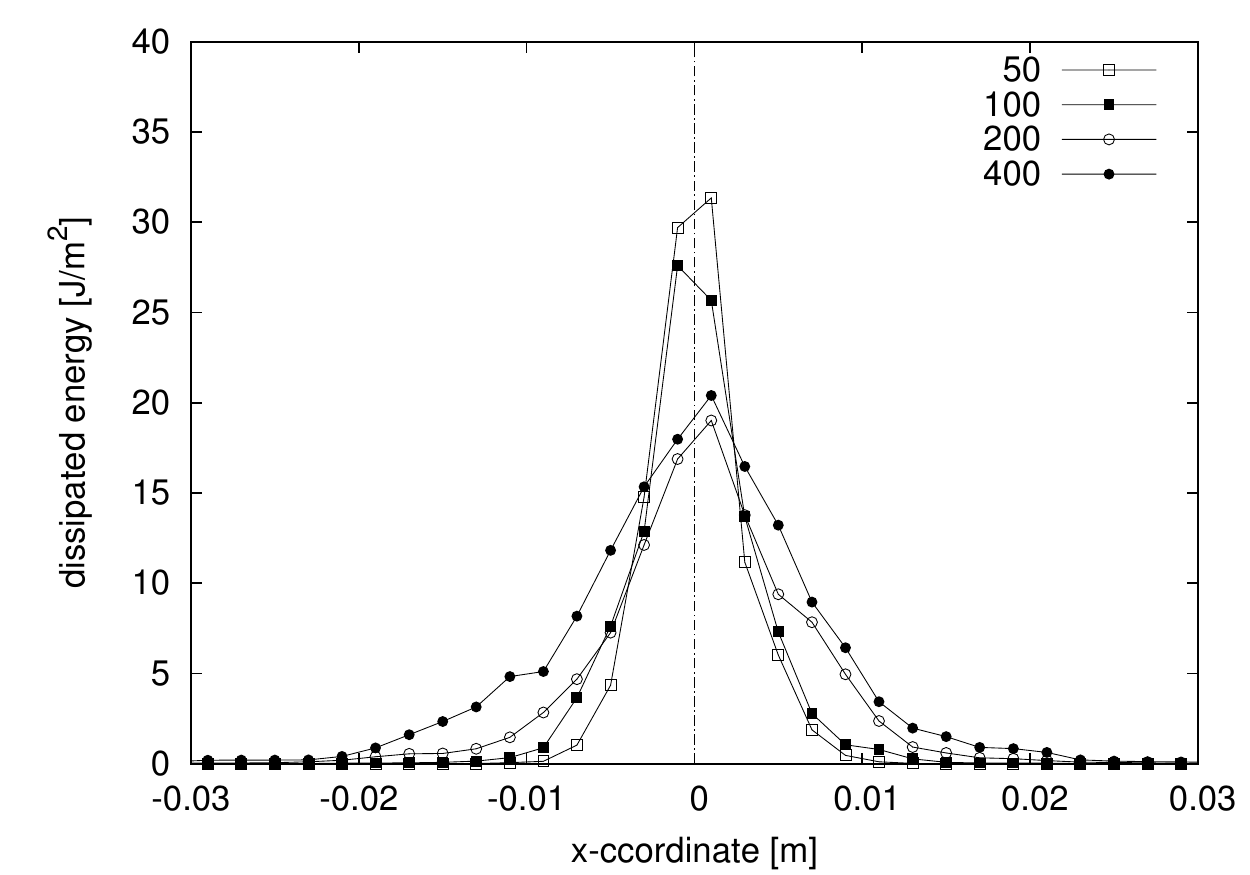}
\end{center}
\caption{Dissipated energy density increment perpendicular to the ligament (x-direction) for the beams with short notch ($a = 0.2$) and four sizes $d = 50$, $100$, $200$ and $400$~mm for a CMOD increment just after the peak of the mean load-CMOD curves. The results are based on an average of 100 random analyses.}
\label{fig:rCurveChangeOppCompSN}
\end{figure}

The analogue distributions for the short notch specimens ($a=0.2$) are depicted in Figs.~\ref{fig:rCurveChangeCompSN}~and~\ref{fig:rCurveChangeOppCompSN}.
Very similar trends as for the long notch specimens are observed. 
However, the fracture process zones are longer in the y-direction than in the long notched specimens.
Nevertheless, the results for long and short notched beams of the same beam depths are difficult to compare, since the ligament length are different. Therefore, the stress gradient, which influences strongly the fracture process, is different as well. 


\begin{figure}
\begin{center}
\includegraphics[width=10cm]{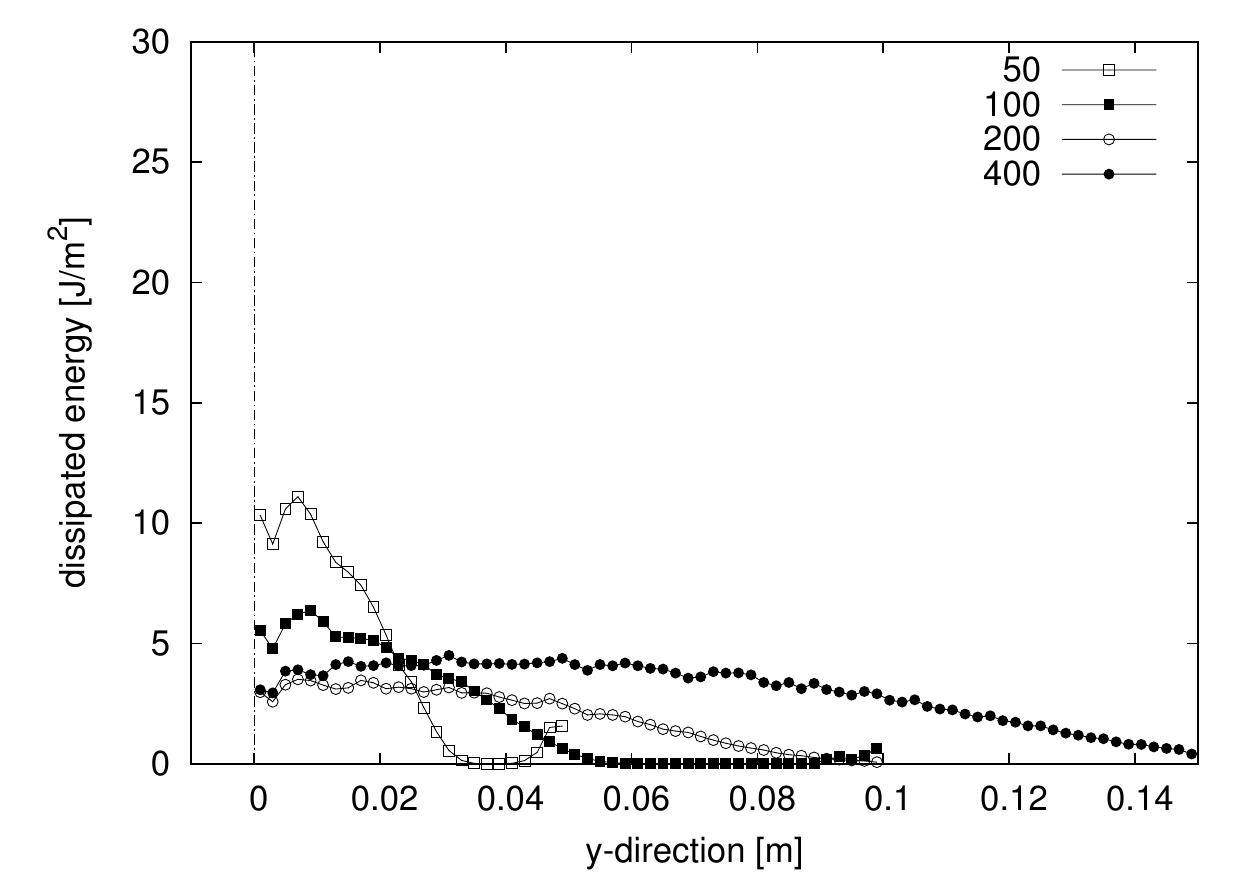}
\end{center}
\caption{Dissipated energy density increment along the ligament (y-direction) for the unnotched beam ($a = 0$) and four sizes $d = 50$, $100$, $200$ and $400$~mm for a CMOD increment just after the peak of the mean load-CMOD curves. The results are based on an average of 100 random analyses.}
\label{fig:rCurveChangeCompUN}
\end{figure}

\begin{figure}
\begin{center}
\includegraphics[width=10cm]{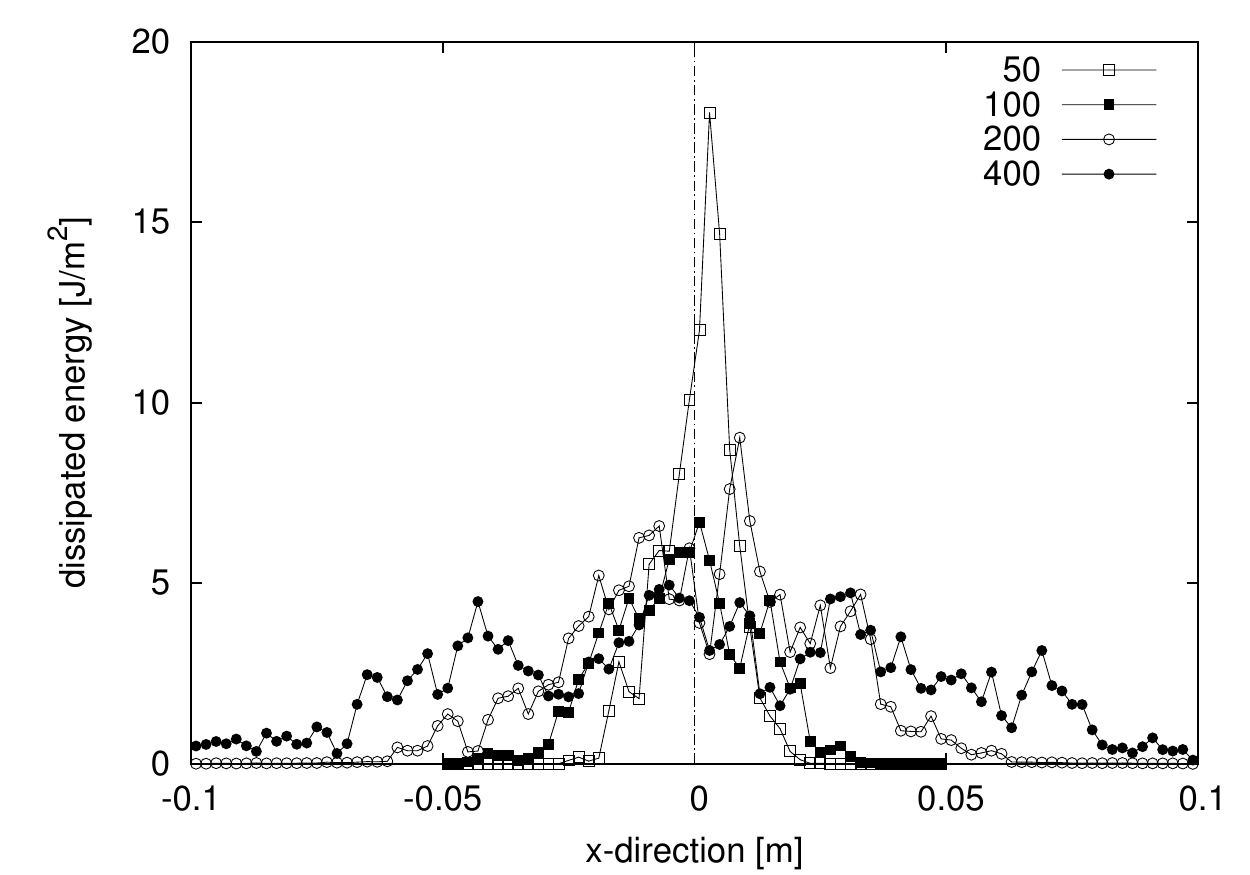}
\end{center}
\caption{Dissipated energy density increment perpendicular to the ligament (x-direction) for the unnotched beam ($a = 0$) and four sizes $d = 50$, $100$, $200$ and $400$~mm for a CMOD increment just after the peak of the mean load-CMOD curves. The results are based on an average of 100 random analyses.}
\label{fig:rCurveChangeOppCompUN}
\end{figure}

\begin{figure}
\begin{center}
\includegraphics[width=10cm]{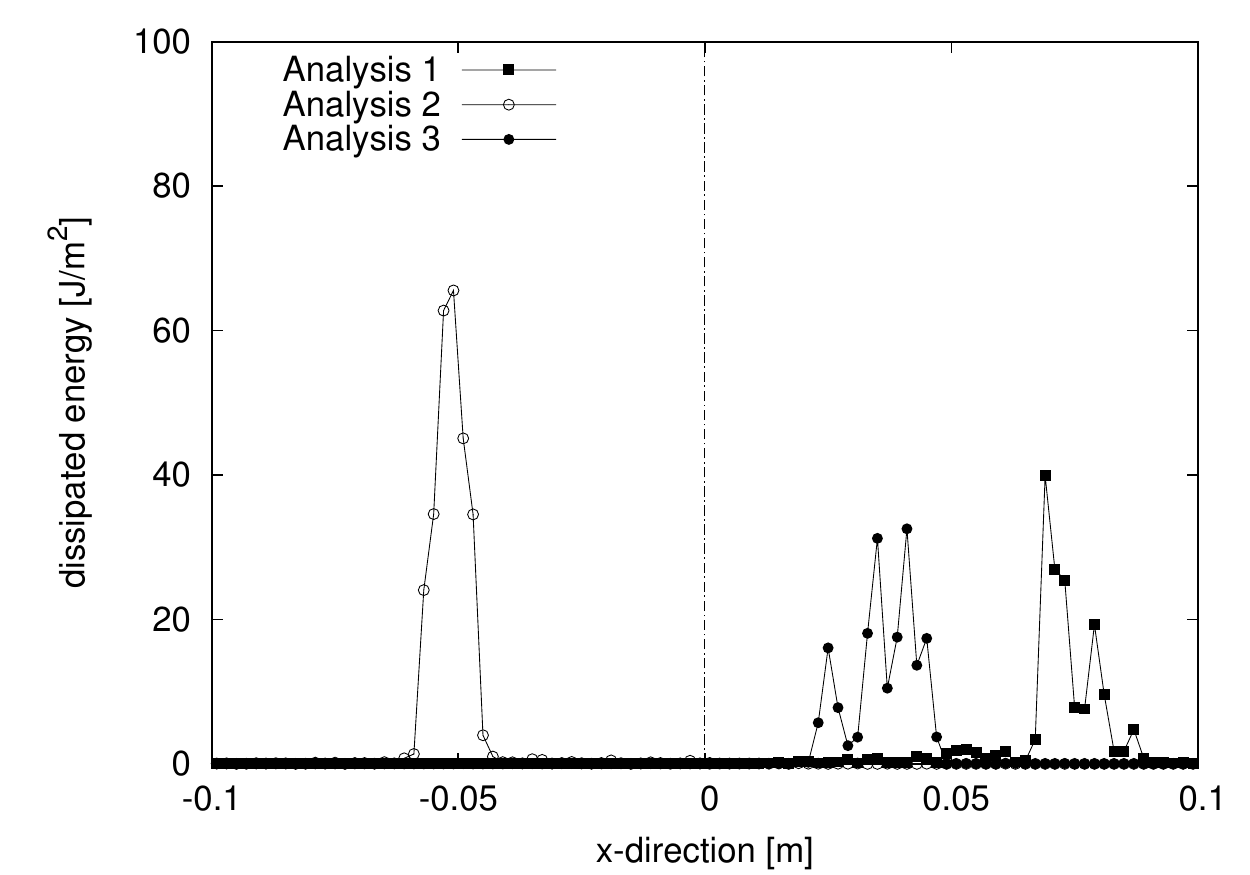}
\end{center}
\caption{Dissipated energy density increment perpendicular to the ligament (x-direction) for three random analyses for the unnotched beam ($a = 0$) and $d = 400$~mm for a CMOD increment just after the peak of the mean load-CMOD curves.}
\label{fig:rCurveChangeOppAnalUN}
\end{figure}

The third set of distributions are shown in Figs.~\ref{fig:rCurveChangeCompUN}~and~\ref{fig:rCurveChangeOppCompUN}.
Again, the beam size influences strongly the length of the fracture process zone. 
The greater the beam size, the longer is the fracture process zone. 
In contrary to the notched specimens, the distribution perpendicular to the ligament is strongly influenced by the beam size. 
For the smallest beam, the width of the fracture process zone is around 40~mm. 
For the largest beam ($d = 400$~mm), the width increases to around $200$~mm. 
This very large width of the fracture process results from the averaging of the densities of 100 random analyses.
Each random analysis is characterised by a localised fracture process zone similar to the profiles shown in Figs.~\ref{fig:dissChange1} for the notched specimens.
For these specimens the notch determines the horizontal centre (x = 0) of the fracture process zone.
On the other hand, for the unnotched specimen, the start of the fracture process zone is determined by the interplay of the bending moment distribution and the random properties of the material.
With increasing beam size, the gradient of the bending moment distribution along the beam decreases, which increases the zone in which the fracture process zone can be initiated.
To illustrate this phenomenon, the dissipated energy densities for three random analyses is shown in Fig.~\ref{fig:rCurveChangeOppAnalUN}.
It could be argued that one should shift the fracture process zones of individual analyses so that they overlap in the centre of the beam, as in \cite{GraJir10} for determining the fracture process zone in a periodic cell subjected to uniaxial tension. However, since in the present three-point bending analyses a stress gradient is present, a shift of the fracture process zone is not required.

\begin{figure}
\begin{center}
\includegraphics[width=10cm]{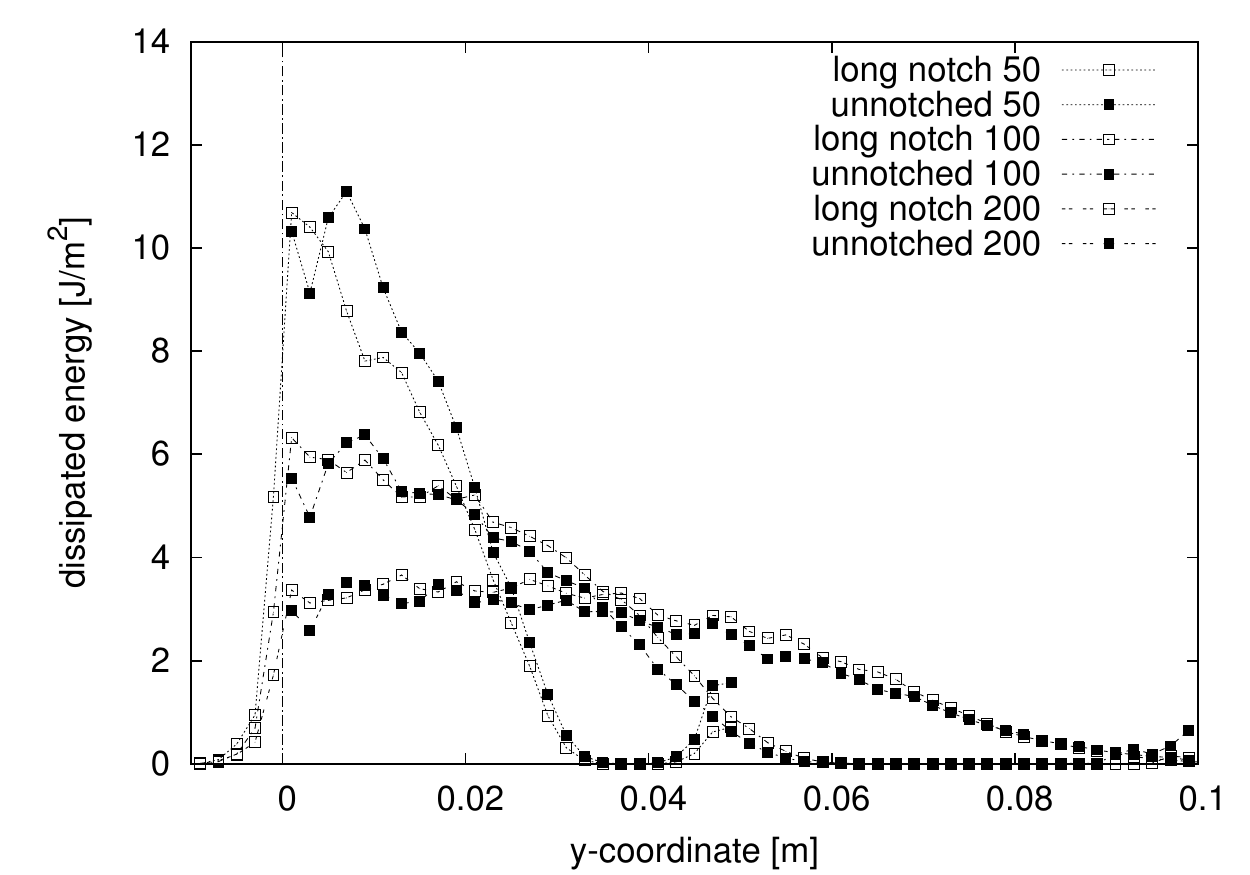}
\end{center}
\caption{Comparison of dissipated energy density increment along the ligament (y-direction) for the beams with and without notch for the same ligament length of $50$, $100$ and $200$~mm for a CMOD increment just after the peak of the mean load-CMOD curve. The results are based on an average of 100 random analyses.}
\label{fig:rCurveLigComp}
\end{figure}

The results for different notch types and the same beam depths cannot be compared directly, as the ligament length differs.
Therefore, the results of the long notched and unnotched specimens with the same ligament length are compared in Fig.~\ref{fig:rCurveLigComp}.
There is almost no influence of the notch type on the distribution of dissipated energy.

\section{Conclusions}

Random meso-scale analyses of three point bending tests were performed to investigate the influence of the size effect on fracture process zones, which were determined from an average of increments of energy just after peak load. 
The study resulted in the following conclusions:

\begin{enumerate}
\item The width of the fracture process zone does not depend on the size of notched specimen. However, for the unnotched specimens, the greater the specimen size, the wider is the fracture process zone. 

\item The length of the fracture process zone along the ligament depends on the size of the specimens. The greater the specimen size, the longer is the region of dissipated energy. However, the length increases nonproportionally to the size.

\item There is no influence of the boundary type on the distribution of the dissipated energy along the ligament for beams with the same ligament length. However, the width of the fracture process zone differs for notched and unnotched beams of the same ligament length.
\end{enumerate}

The modelling techniques used in the present study are based on many simplifications stated in the introduction of this work. Therefore, additional experiments, which provide local information on fracture processes, should be carried out to verify the local results presented here.

\section*{Acknowledgement}

Partial financial support by the ERC advanced grant Failflow (27769) is gratefully acknowledged.
The meso-scale analyses were performed with the finite element package OOFEM \citep{Pat99,PatBit01} extended by the present authors.

\bibliographystyle{plainnat}
\bibliography{part}

\end{document}